%% file: pvis-tracking-revision-main.tex
\documentclass[journal]{vgtc}                     

\ifXeTeX
  \DeclareGraphicsExtensions{.pdf,.png,.jpg,.jpeg}
  \usepackage{newunicodechar}
  \newunicodechar{—}{\textemdash}
  \newunicodechar{–}{\textendash}
  \newunicodechar{’}{\textquoteright}
\fi

\vgtccategory{Research}

\graphicspath{{./figs/}} 
\usepackage{booktabs}
\usepackage{amsmath}
\usepackage{amsfonts}
\usepackage{amssymb}

\usepackage{bm}

\usepackage{tabularx}
\usepackage[dvipsnames,svgnames]{xcolor}

\usepackage{multirow}
\usepackage{makecell}

\newcommand{\para}[1]{\noindent{\textbf{#1}}}

\title{Direct Topology Tracking in Continuous Implicit Models}

\author{
  \authororcid{Guanqun Ma}{0000-0001-8102-3172}, 
  \authororcid{David Lenz}{0000-0002-2587-2783},
  \authororcid{Kaiyuan Tang}{0009-0001-3512-0112},
  \authororcid{Hanqi Guo}{0000-0001-7776-1834},
  \authororcid{Chaoli Wang}{0000-0002-0859-3619},
  \authororcid{Tom Peterka}{0000-0002-0525-3205}, 
  and \authororcid{Bei Wang}{0000-0002-9240-0700}
}
\authorfooter{
\item Guanqun Ma and Bei Wang are with the University of Utah.  \\
E-mail: guanqun.ma@utah.edu; beiwang@sci.utah.edu.

\item David Lenz and Tom Peterka are with Argonne National Laboratory.  \\
E-mail: \{dlenz, tpeterka\}@anl.gov.

\item Kaiyuan Tang and Chaoli Wang are with the University of Notre Dame.  \\
E-mail: \{ktang2, chaoli.wang\}@nd.edu.

\item Hanqi Guo is with The Ohio State University.
E-mail: guo.2154@osu.edu.
}

\abstract{
\input{sec-abstract}

}

\keywords{Feature tracking, continuous implicit models, multivariate functional approximations, implicit neural representations, topological data analysis}

\teaser{
\centering
\includegraphics[width=0.95\linewidth]{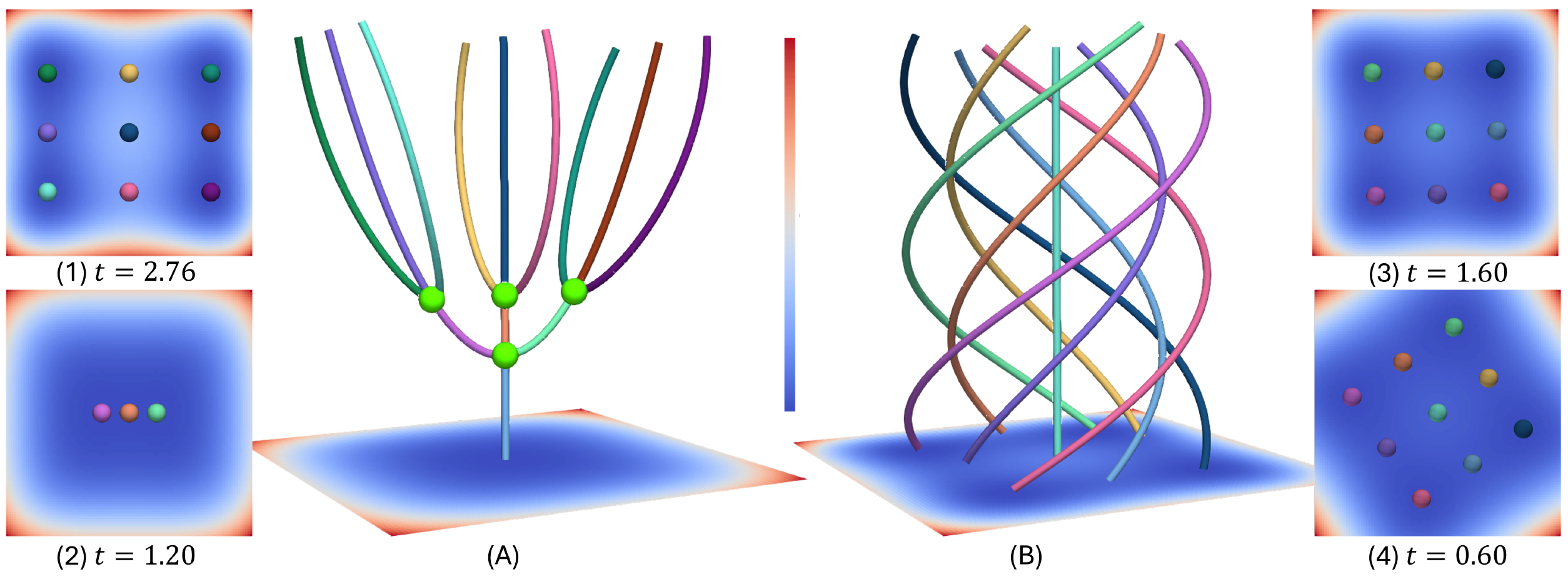}
\vspace{-2mm}
\caption{Tracking results for (A) the Quartic Potential model and (B) the Quartic Rotation model. Bright green spheres indicate degenerate points. Panels (1, 2) and (3, 4) present representative snapshots of the scalar fields for (A) and (B), respectively.}
\label{fig:teaser}
}

\begin{document}

\maketitle

\input{sec-introduction}

\input{sec-related-work}
\input{sec-background}

\input{sec-method}

\input{sec-configuration}

\input{sec-analytic-results}
\input{sec-scientific-results}

\input{sec-conclusion}
\input{sec-ack}
\bibliographystyle{abbrv-doi-hyperref-narrow}
\bibliography{refs-tracking}       
\clearpage
\newpage
\appendix
\input{sec-app-step-size-hardware}
\input{sec-app-finite-difference}

\input{sec-app-complexity}
\input{sec-app-analytic-parameters}
\input{sec-app-scientific-parameters}
\input{sec-app-necessity-of-newton-correction}

\input{sec-app-necessity-of-degenerate-cases}
\input{sec-app-analytic-validation}
\input{sec-app-real-data-reference}
\input{sec-app-additional-results}


\end{document}

%% file: sec-abstract.tex
We present a framework for tracking topological features directly within continuous implicit models. Such models, including implicit neural representations (INRs) and multivariate functional approximations (MFAs), are increasingly adopted to represent scientific data without the resolution constraints of discrete grids. They offer compact, smooth, and differentiable representations of complex fields, enabling new opportunities for high-performance data storage, reconstruction, and analysis. Given a continuous implicit model, our method tracks the evolution of critical points by querying the model and its derivatives, thereby eliminating the need to resample onto a grid. This approach enables faithful feature tracking while avoiding discretization-induced artifacts such as aliasing. We demonstrate the generality of our framework across a range of implicit representations, including analytic functions, MFAs, and INRs, and show that it produces smooth, coherent critical point trajectories. By enabling feature tracking directly on continuous representations, our method supports a new class of feature-driven visualization workflows centered on implicit models.

%% file: sec-introduction.tex
\label{sec:introduction}
\section{Introduction}

As scientific simulations continue to grow in scale and fidelity, continuous implicit models such as multivariate functional approximations (MFAs) and implicit neural representations (INRs)~\cite{majdisova2017radial,sitzmann2020implicit,mildenhall2021nerf} are increasingly used to represent scientific data, offering promising opportunities for next-generation scientific workflows. A continuous implicit model represents a field as a function whose value at any point in space and/or time is obtained by evaluating the function, rather than by storing samples on a discrete grid. The field is therefore encoded \emph{implicitly} in the function and defined \emph{continuously} across its entire domain. These models are transforming scientific workflows~\cite{niemeyer2021giraffe,sitzmann2019scene} by providing compact data representations, fast surrogate modeling, and reduced I/O overhead in large-scale simulations, while supporting differentiable, resolution-independent queries, alias-free continuous analysis, and improved exploration and visualization. 

MFAs represent fields using smooth tensor-product B-splines, enabling direct evaluation of field values and higher-order derivatives~\cite{peterka2018foundations,peterka2022multivariate}.
INRs use neural networks to learn coordinate-based representations from data. SIREN employs sinusoidal activations to represent signals and their derivatives~\cite{sitzmann2020implicit}, while CoordNet targets time-varying scientific data~\cite{han2023coordnet}. These models enable direct analysis of scientific fields through their functional representations.

However, their potential for topological data analysis remains largely unexplored. Topological descriptors have long provided powerful tools in scientific visualization, supporting tasks such as feature detection and tracking~\cite{heine2016survey,yan2021scalar}. Yet despite this promise, most existing techniques operate on discretized data and require sampling onto a grid before extracting topological features. When applied to continuous implicit models, such discretization discards their inherent smoothness and differentiability and may introduce aliasing or resolution-dependent artifacts.
This gap raises a central question: can topological features be extracted and tracked directly in continuous implicit models, without discretizing the domain? Recent progress suggests that the answer is within reach. In terms of feature extraction, Ma et al.~\cite{ma2024critical} extracted critical points directly from MFAs, and subsequent work enabled the extraction of contours, Jacobi sets, and ridge-valley graphs~\cite{ma2025extracting}. Building on these developments, we investigate how to track critical points directly in time-varying continuous implicit models.

Feature tracking reveals how structures appear, disappear, merge, and split. Such analysis is critical in domains such as climate~\cite{engelke2021topology}, combustion~\cite{bremer2011interactive}, and cosmology~\cite{friesen2016situ}, where temporal behavior is often as important as spatial structure. 
Feature Flow Fields (FFF)~\cite{theisel2003feature} provide an established principle for continuation by representing feature trajectories as streamlines of a derived space-time vector field. Inspired by this principle, we develop a tracking pipeline for critical points of continuous implicit models.

While FFF provides a continuation principle for tracing feature trajectories, constructing a graph of critical-point trajectories also requires controlling numerical drift~\cite{weinkauf2011stable} and resolving branch connectivity at degenerate events. Numerical integration can drift away from the zero-gradient condition that defines a critical point. Furthermore, tracing an individual trajectory does not by itself identify all branches incident to a degenerate event or determine how they connect.

We address these challenges by integrating FFF-based continuation with constraint correction and explicit event reconstruction into a comprehensive framework for tracking critical points directly in continuous implicit models. Our contributions are:

\begin{itemize}[noitemsep,leftmargin=*]
\item \textbf{An integrated tracking framework for continuous implicit models.}
We combine seed identification, FFF-based continuation, trajectory correction, event reconstruction, and graph construction using direct queries of the model and its derivatives, without constructing a sampled surrogate.

\item \textbf{Constraint correction and explicit degenerate-event reconstruction.}
We combine ODE prediction with fixed-time Newton correction to reduce drift from the zero-gradient condition. We also explicitly localize isolated degenerate events and reconstruct their incident branches and connectivity, including configurations in which multiple branches merge or split.

\item \textbf{Evaluation across implicit representations.}
We evaluate the complete workflow on 2D and 3D time-varying analytic, MFA, and INR models, demonstrating direct tracking across these representations without constructing a full discrete grid.
\end{itemize}

%% file: sec-related-work.tex
\section{Related Work}
\label{sec:related-work}

\subsection{Continuous Implicit Models}

Continuous implicit models provide smooth and differentiable representations of scalar fields~\cite{rella2024neural,bloomenthal1997introduction}. Because they are not bound to a fixed-resolution grid, these models can be evaluated at arbitrary locations within the domain. This flexibility enables direct access to both function values and derivatives, which is valuable for scientific data analysis and visualization~\cite{gomes2009implicit,ma2024critical}.

\para{Multivariate functional approximation} (MFA) constructs smooth B-spline functions for multivariate scientific data~\cite{peterka2018foundations,peterka2022multivariate,lenz2023customizable}. MFA belongs to the broader family of scattered data approximation (SDA) methods~\cite{wendland2004scattered}, in which a continuous model is built directly from sample values. Numerous SDA techniques exist, including spline-based~\cite{deBoor2001guide}, wavelet-based~\cite{jansen2005second}, and radial basis function approaches~\cite{majdisova2017radial}. MFA is particularly attractive because B-splines offer high continuity, compact parameterization, and efficient evaluation of function values and higher-order derivatives throughout the domain~\cite{peterka2018foundations}. These properties make MFA well-suited for simulation analysis, feature extraction, and visualization tasks where smoothness and differentiability are essential.

Supported by the U.S. Department of Energy (DOE) through the SciDAC RAPIDS Institute~\cite{RAPIDS2}, MFA has demonstrated its utility in high-energy physics~\cite{HepOnHPC}, climate modeling~\cite{Seahorce}, and large-scale volume rendering~\cite{sun2023scalable}, establishing it as a practical representation for high-performance computing workflows. Ma et al.~\cite{ma2024critical} extracted critical points directly from MFA, showing that topological features can be computed without discretizing the model. Because critical points underpin higher-level topological structures—such as Morse-Smale complexes and critical point trajectories in time-varying data—this result positions MFA as a promising substrate for topological analysis of continuous implicit models. Most recently, Ma et al.~\cite{ma2025extracting} extended MFA-based topology extraction to contours, Jacobi sets, and ridge-valley graphs, further demonstrating the potential of MFA to support a broad range of topological descriptors. Together, these developments suggest that MFA is emerging as a robust foundation for future topology-aware analysis and visualization pipelines.

\para{Implicit neural representations (INRs)} represent spatial or time-varying data by learning mappings from coordinates to signal values. Sitzmann et al.~\cite{sitzmann2020implicit} introduced SIREN, which models signals using sinusoidal activations. For scientific volumes, INR approaches support compact encoding~\cite{lu2021compressive}, fast rendering~\cite{weiss2022fast}, and efficient representation using adaptive feature grids~\cite{wurster2024adaptively} and hash encoding~\cite{devkota2023efficient}. Related developments include meta-learning-based encoding~\cite{yang2025meta}, distributed representations~\cite{wu2025distributed}, and parameter exploration~\cite{chen2025explorable}.

INRs for time-varying scientific data are particularly relevant to critical-point tracking. Han and Wang~\cite{han2023coordnet} introduced CoordNet, which maps coordinates to signal values using residual blocks and periodic activations. Their framework supports data and visualization generation for time-varying volumes, including spatial and temporal super-resolution. Tang and Wang~\cite{tang2024ECNR} proposed ECNR, which combines a Laplacian pyramid with multiple small neural networks to encode local content and residual blocks across scales for time-varying volume compression. Han et al.~\cite{han2026moe} introduced MoE-INR, which uses a policy network to route spatiotemporal coordinates to expert decoders that share an encoder. These methods advance the representation, reconstruction, and compression of evolving scientific fields.

We investigate critical-point tracking directly on continuous neural fields by querying field values and the spatial and temporal derivatives needed for continuation and event localization. Our INR experiments use CoordNet~\cite{han2023coordnet} as a representative model for time-varying scientific data. Applicability to other architectures depends on their smoothness and support for the required derivative queries.

\subsection{Feature Tracking}

Feature tracking is a longstanding topic in scientific visualization, and topological descriptors have proven particularly effective for summarizing and comparing features~\cite{yan2021scalar}. Reininghaus et al.~\cite{reininghaus2011efficient} introduced a combinatorial formulation for tracking critical points in time-dependent scalar fields by combining the combinatorial gradient field~\cite{forman1998combinatorial} with persistence~\cite{edelsbrunner2001hierarchical}. Saikia and Weinkauf~\cite{saikia2017global} extended feature tracking beyond individual critical points to merge-tree-based regions, helping automate the challenging task of identifying meaningful similarities in large datasets. Persistence diagrams equipped with Wasserstein-distance extensions have also been explored for topology tracking~\cite{soler2018lifted,soler2019ranking}. Pont et al.~\cite{pont2022wasserstein} introduced a merge-tree metric based on $L^2$-Wasserstein distances between extremum persistence diagrams. Yan et al.~\cite{yan2023geometry} incorporated geometric information through labeled interleaving distances to enable more accurate matching of critical points, while Li et al.~\cite{li2025flexible} employed partial optimal transport to support flexible and partial feature matching.

\para{Relation to (Stable) Feature Flow Fields.}
Most closely related to our work, Theisel and Seidel introduced Feature Flow Fields (FFF)~\cite{theisel2003feature}, which represent feature trajectories as streamlines of a derived space-time vector field. FFF established a derivative-based alternative to matching features independently across time steps and demonstrated this idea for critical points in time-dependent vector fields.

Our work builds on this continuation principle rather than introducing a new feature-velocity formulation. In FFF, a feature trajectory is represented as an integral curve of a derived vector field. Consequently, a numerical integration that drifts away from the feature line does not explicitly re-enforce the defining feature constraint. In contrast, we formulate critical-point continuation directly from the zero-gradient condition of a time-varying scalar field represented by a continuous implicit model. The model and its derivatives are queried directly, avoiding construction of a sampled vector-field surrogate. 
After each ODE prediction, we explicitly refine the spatial position at fixed time until the zero-gradient condition is satisfied to the prescribed tolerance, correcting numerical drift on regular trajectory segments. We further handle isolated degenerate events, where the Hessian becomes singular and trajectories may merge, split, appear, or disappear, by locally reconstructing and connecting their incident trajectory branches.

Stable Feature Flow Fields (SFFF)~\cite{weinkauf2011stable} improve the numerical stability of FFF by modifying the auxiliary flow so that feature lines are locally attracting. The published SFFF formulation for critical-point tracking is limited to 2D time-dependent vector fields. Its 3D application addresses parallel-vector lines rather than time-varying critical points. In contrast, our formulation is expressed for a $d$-dimensional scalar field and is evaluated on both 2D and 3D time-varying models. The same formulation extends to higher dimensions when the required derivatives are available, although the computational cost increases with dimension.  
In the critical-point formulations of FFF and SFFF, generic pair creation or annihilation occurs at a fold bifurcation, where a feature trajectory is tangent to a constant-time slice. Our method explicitly localizes degenerate points and reconstructs the trajectory branches that meet at each point. The resulting trajectory graph captures degenerate events, including those involving multiple merging or splitting branches.

%% file: sec-background.tex
\section{Technical Background}
\label{sec:technical-background}

We briefly review the continuous models used in our framework, MFA and INR, as well as particle tracing.

\subsection{Multivariate Functional Approximation}
\label{sec:MFA-concept}

MFA models represent data using smooth tensor-product B-splines. Rather than storing samples directly on a grid, MFA encodes the underlying field as a continuous function whose coefficients are compactly represented via control points. This representation supports analytic differentiation, multi-resolution evaluation, and continuous queries in arbitrary dimensions. We provide a brief overview here and refer to~\cite{deBoor2001guide, piegl1997nurbs} for comprehensive treatments of B-splines.

A degree-$p$ B-spline is a piecewise-polynomial curve with $p-1$ continuous derivatives. Its domain is partitioned into knot spans by a knot sequence, and its shape is determined by a set of control points. In a nutshell, a knot sequence is an ordered, non-decreasing list of parameter values $\{t_0, t_1, \dots\}$ that specifies where and how the B-spline basis functions join, whereas a knot span is the interval between two consecutive knots,
$[t_i, t_{i+1}]$. Control points are the points that shape the spline curve or surface; they act as weighted anchors that pull the curve or surface toward them, even though the spline does not generally pass through the control points.

\begin{figure}[!ht]
\vspace{-2mm}
\centering
\includegraphics[width=.8\linewidth]{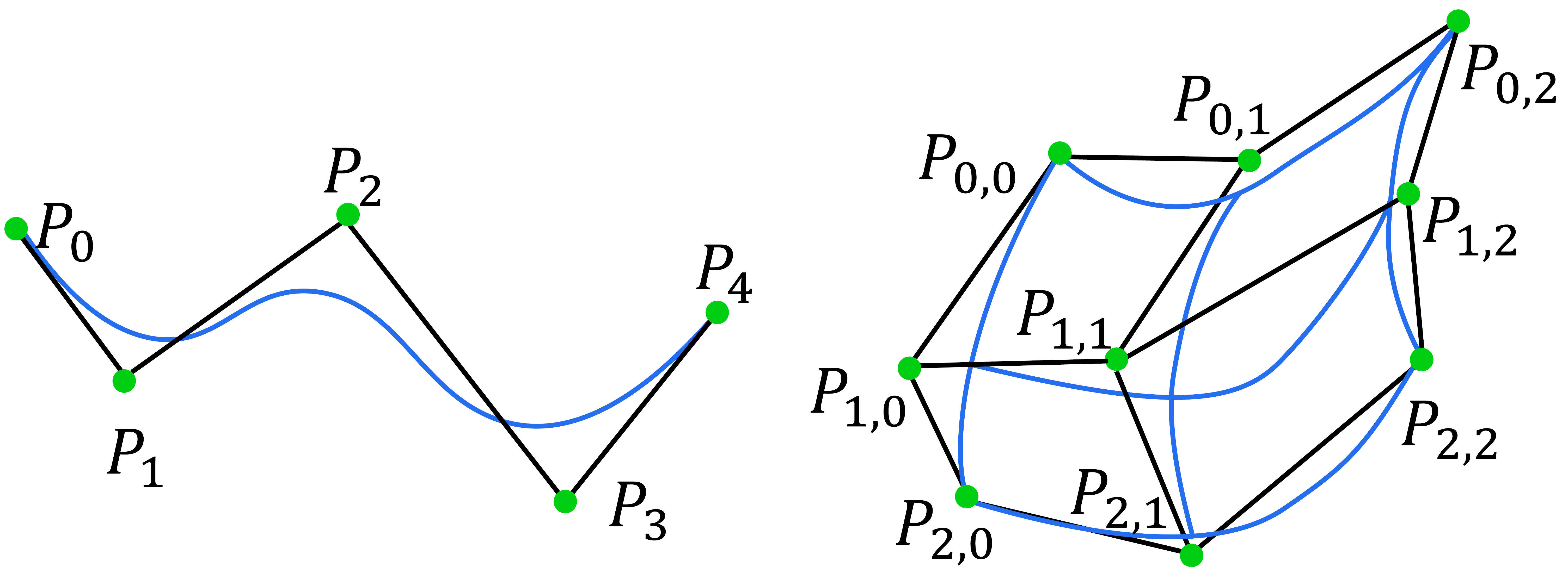}
\vspace{-2mm}
\caption{Left: a 1D B-spline curve. Right: a 2D B-spline surface.
Control points $P_i$ and $P_{i,j}$ are shown in green, the control mesh in black, and the resulting curve or surface in blue.}
\label{fig:mfa-1d-2d}
\vspace{-2mm}
\end{figure}

In the 1D case, let $N_j(u)$ denote the basis functions associated with control points $P_j$. The resulting function is expressed as
\begin{equation}
    F(u) = \sum_{j=0}^{n-1} N_j(u) P_j.
\end{equation} 
Given samples $\{(u_i, f_i)\}_{i=0}^{m-1}, u_i \in [0,1]$, the optimal B-spline control points are obtained through least-squares minimization,
\begin{equation}
    \min_P \left(\frac{1}{m}\sum_{i=0}^{m-1} |f_i - \sum_{j=0}^{n-1} N_j(u_i) P_j|^2 \right)^{1/2}.
\end{equation}
This 1D formulation generalizes directly to multidimensional domains through tensor-product basis construction. 
\cref{fig:mfa-1d-2d} shows a 1D B-spline curve and a 2D B-spline surface.

For the general $d$-dimensional case, the B-spline is
\begin{equation}
  \begin{aligned}
      F(u_1,&\cdots, u_d)=\\
      &\sum_{j_1}\cdots \sum_{j_d} N^{(1)}_{j_1, p}(u_1)\cdots N^{(d)}_{j_d, p}(u_d) P_{j_1,\cdots, j_d}.
  \end{aligned}
  \label{eq:MFA}  
\end{equation}

\subsection{Implicit Neural Representation: CoordNet}
\label{sec:CoordNet}

\begin{figure}[!ht]
\centering
\vspace{-3mm}
\includegraphics[width=\linewidth]{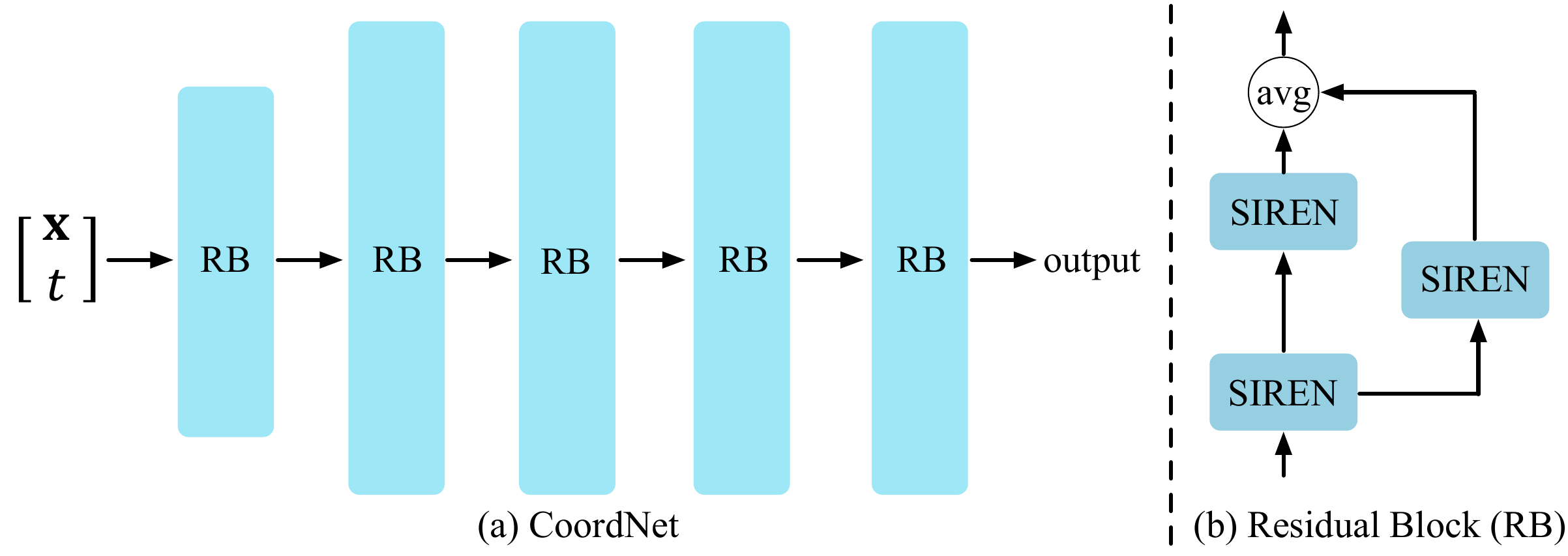}
\vspace{-6mm}
\caption{(a) The architecture of CoordNet, where RB denotes a residual block. (b) The detailed structure of the SIREN-based residual block used in CoordNet.}
\label{fig:CoordNet-overview}
\vspace{-3mm}
\end{figure}

CoordNet~\cite{han2023coordnet} is a representative coordinate-based INR designed for modeling scientific data.
As illustrated in \cref{fig:CoordNet-overview}, it takes spatial-temporal coordinates $(\mathbf{x}, t)$ as input and predicts the corresponding data values. Once trained, the optimized CoordNet model provides a continuous implicit representation of the underlying field, enabling value queries at arbitrary spatial-temporal locations.

The architecture of CoordNet is fully connected and composed of SIREN-based residual blocks, where each block contains SIREN layers with sinusoidal activations~\cite{sitzmann2020implicit}. These residual blocks stabilize training, support deeper network structures, and improve reconstruction fidelity, while the sinusoidal activations enhance the model’s ability to capture fine-scale structures and high-frequency variations commonly found in scientific data.

In this paper, we use a CoordNet model with five residual blocks for each dataset to achieve high-fidelity representation and efficient feed-forward evaluation. Aside from the number of residual blocks, all training hyperparameters follow the default CoordNet configuration. Because our critical point tracking framework requires high numerical precision, we convert all learned parameters from \texttt{float32} to \texttt{float64} at inference time and evaluate the optimized CoordNet model in double precision, consistent with MFA, to ensure sufficient numerical accuracy for subsequent processing.

\subsection{Particle Tracing}
\label{sec:background-particle-tracing}

Particle tracing follows virtual particles advected by a vector field and is widely used to characterize flow behavior through trajectory-based representations~\cite{peterka2011study}. The motion of a particle in a time-dependent velocity field is described by the following initial value problem (IVP)~\cite{ma2025extracting}: 
\begin{equation}
\label{eq:initial-value-problem}
\frac{d\mathbf{x}}{dt} = \mathbf{v}(t,\mathbf{x}),
\qquad
\mathbf{x}(t_0) = \mathbf{x}_0,
\end{equation}
where $\mathbf{x}(t)$ denotes the particle’s position at time $t$, and $\mathbf{v}(t,\mathbf{x})$ is the velocity at $(t,\mathbf{x})$.
Particle trajectories are obtained by numerically integrating this IVP~\cite{pokrajac2002efficient}.

A widely used integrator is the classical fourth-order Runge–Kutta method (RK4)\cite{press2007numerical}.
Given a current state $\mathbf{x}_n$ at time $t_n$, RK4 obtains the solution to $(t_{n+1}, \mathbf{x}_{n+1})$ using

\begin{equation}
\label{eq:RK4}
    \begin{aligned}
        \mathbf{x}_{n+1} &= \mathbf{x}_n + \frac{s_t}{6} \left( \mathbf{w}_1 + 2\mathbf{w}_2 + 2\mathbf{w}_3 + \mathbf{w}_4 \right), \\
t_{n+1} &= t_n + s_t,
    \end{aligned}
\end{equation}
with intermediate slopes defined by
$\mathbf{w}_1 = \mathbf{v}(t_n, \mathbf{x}_n)$,
$\mathbf{w}_2 = \mathbf{v}\!\left(t_n + \tfrac{s_t}{2},\, \mathbf{x}_n + \tfrac{s_t}{2}\mathbf{w}_1\right)$,
$\mathbf{w}_3 = \mathbf{v}\!\left(t_n + \tfrac{s_t}{2},\, \mathbf{x}_n + \tfrac{s_t}{2}\mathbf{w}_2\right)$,
$\mathbf{w}_4 = \mathbf{v}(t_n + s_t,\, \mathbf{x}_n + s_t\mathbf{w}_3)$.
Here, $s_t>0$ is the temporal step size.
The method produces an approximation to $\mathbf{x}(t_{n+1})$ with a local truncation error of $O(s_t^5)$ and an accumulated error of $O(s_t^4)$\cite{press2007numerical}.

\cref{eq:RK4} provides the time-parameterized RK4 update.
For applications requiring spatially stable integration, we adopt the following arc-length-parameterized RK4 scheme \cite{press2007numerical}:
\begin{equation}
\begin{aligned}
        &\frac{d\mathbf{x}}{dr}=\frac{\mathbf{v}(t(r),\, \mathbf{x}(r))}{\left\|\mathbf{v}(t(r),\, \mathbf{x}(r))\right\|} = \mathbf{v}^\mathbf{x}, \quad
        \frac{dt}{dr}=\frac{1}{\left\|\mathbf{v}(t(r),\, \mathbf{x}(r))\right\|} = v^t, \\
        &\mathbf{x}(0)=\mathbf{x}_0,\quad t(0) = t_0.
\end{aligned}
\end{equation}
RK4 updates $(t_{n+1}, \mathbf{x}_{n+1})$ in
\begin{equation}
\label{eq:RK4-spatial}
    \begin{aligned}
        \mathbf{x}_{n+1} &= \mathbf{x}_n + \frac{s_s}{6} \left( \mathbf{w}^\mathbf{x}_1 + 2\mathbf{w}^\mathbf{x}_2 + 2\mathbf{w}^\mathbf{x}_3 + \mathbf{w}^\mathbf{x}_4 \right), \\
t_{n+1} &= t_n + \frac{s_s}{6} \left(w^t_1 + 2w^t_2 + 2w^t_3 + w^t_4 \right),
    \end{aligned}
\end{equation}
with intermediate slopes defined by
\begin{equation}
    \begin{aligned}
        &\mathbf{w}^\mathbf{x}_1 = \mathbf{v}^\mathbf{x}(t_n, \mathbf{x}_n),\quad w_1^t = v^t(t_n, \mathbf{x}_n),\\
        &\mathbf{w}^\mathbf{x}_2 = \mathbf{v}^\mathbf{x}\left(t_n + \tfrac{s_s}{2}w_1^t, \mathbf{x}_n + \tfrac{s_s}{2}\mathbf{w}^\mathbf{x}_1\right), \\
        & w^t_2 = v^t\!\left(t_n + \tfrac{s_s}{2}w_1^t,\, \mathbf{x}_n + \tfrac{s_s}{2}\mathbf{w}^\mathbf{x}_1\right), \\
        &\mathbf{w}^\mathbf{x}_3 = \mathbf{v}^\mathbf{x}\left(t_n + \tfrac{s_s}{2}w^t_2, \mathbf{x}_n + \tfrac{s_s}{2}\mathbf{w}^\mathbf{x}_2\right), \\
        &w^t_3 = v^t\!\left(t_n + \tfrac{s_s}{2}w^t_2,\, \mathbf{x}_n + \tfrac{s_s}{2}\mathbf{w}^\mathbf{x}_2\right),\\
        &\mathbf{w}^\mathbf{x}_4 = \mathbf{v}^\mathbf{x}(t_n + s_sw^t_3,\, \mathbf{x}_n + s_s\mathbf{w}_3^\mathbf{x}) \\
        & w^t_4 = v^t(t_n + s_sw^t_3,\, \mathbf{x}_n + s_s\mathbf{w}^\mathbf{x}_3).
    \end{aligned}
\end{equation}
Here $s_s>0$ is the spatial step size.
In summary, time-parameterized RK4 uses the \emph{temporal step size} $s_t$ to advance the integration in time, with the corresponding spatial displacement determined accordingly. In contrast, arc-length-parameterized RK4 uses the \emph{spatial step size} $s_s$ to control the spatial displacement, with the corresponding temporal change determined accordingly.

%% file: sec-method.tex
\section{Method}
\label{sec:method}

To our knowledge, this work presents the \emph{first} approach for feature tracking---in particular, critical point tracking---directly on continuous implicit models. Although our examples focus on closed-form analytic functions, MFAs, and INRs, the method is general and applies to any implicit representation that supports function evaluation and higher-order derivatives.

\subsection{Critical Point Tracking}
\label{sec:critical-point-tracking}

Let $f : \Omega \times \mathbb{R} \rightarrow \mathbb{R}$ be a smooth scalar field defined on a spatial domain $\Omega \subset \mathbb{R}^d$ and parameterized by time $t \in \mathbb{R}$. 

We denote a spatial position by $\mathbf{x} = (x_0, x_1, \dots, x_{d-1})^\top \in \Omega$. A trajectory of a time-varying critical point is a continuous path $\mathbf{x}(t)$ satisfying 
$\nabla_{\mathbf{x}} f(\mathbf{x}(t), t) = \mathbf{0}$.
The motion of a critical point follows an IVP:
\begin{equation}
\label{eq:ODE}
    \frac{d\mathbf{x}}{dt} = \xi(t,\mathbf{x}),\quad \mathbf{x}(t_0) = \mathbf{x}_0, \quad \nabla_{\mathbf{x}} f(\mathbf{x}_0, t_0) = \mathbf{0},
\end{equation}
where $\xi(t,\mathbf{x})$ denotes the instantaneous velocity of the critical point, determined by differentiating the criticality condition with respect to time. 
For all critical points, as $\nabla_{\mathbf{x}} f \equiv \mathbf{0}$, total derivative of $\nabla_{\mathbf{x}} f$ vanishes:
\begin{equation}
\frac{d}{dt}\,\nabla_{\mathbf{x}} f(\mathbf{x}(t), t)
= \mathrm{H}_{\mathbf{x}} f(\mathbf{x}(t), t)\frac{d\mathbf{x}}{dt}
+ \frac{\partial}{\partial t}\nabla_{\mathbf{x}} f(\mathbf{x}(t), t)
= \mathbf{0},
\end{equation}
where $\mathrm{H}_{\mathbf{x}}f=\nabla_{\mathbf{x}}^2f$ denotes the spatial Hessian of $f$. Solving for the velocity gives
\begin{equation}
\label{eq:velocity}
\xi(t,\mathbf{x})=\frac{d\mathbf{x}}{dt}
= -\left[\mathrm{H}_{\mathbf{x}} f(\mathbf{x}(t), t)\right]^{-1}
\,\frac{\partial}{\partial t}\nabla_{\mathbf{x}} f(\mathbf{x}(t), t).
\end{equation}
The derivation assumes that $\mathrm{H}_{\mathbf{x}} f$ is nonsingular along regular portions of the trajectory. Degenerate cases where this condition fails are handled separately in \cref{sec:degenerate-case}.
A formulation similar to \cref{eq:velocity} has been derived in the context of Feature Flow Fields~\cite{theisel2003feature}. We generalize this formulation to higher dimensions and integrate it into our pipeline.

\begin{figure}[!ht]
\centering
\includegraphics[width=1.04\linewidth]{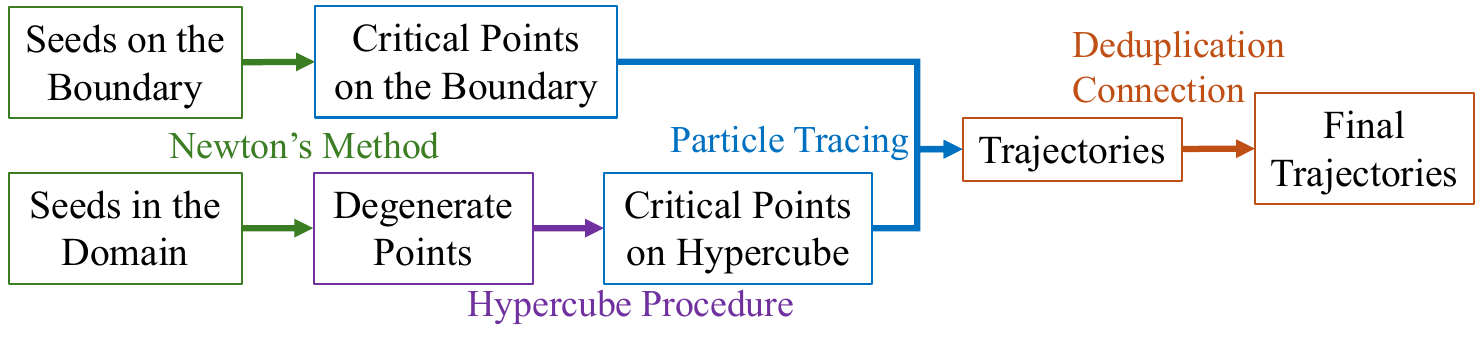}
\vspace{-7mm}
\caption{Flowchart of our critical point tracking pipeline.}
\label{fig:method-flow-chart}
\vspace{-5mm}
\end{figure}

We numerically integrate \cref{eq:ODE} using RK4. Starting from an initial critical point $p_0$, successive RK4 updates produce a discrete trajectory $\{x_0, x_1, x_2, \ldots\}$, which approximates the underlying path.
Our tracking pipeline consists of three major steps, shown in \cref{fig:method-flow-chart}:
\begin{enumerate}[noitemsep,leftmargin=*]
\item \textbf{Initialization:} Identify all critical points on the space-time boundary from which trajectories will be initiated.
\item \textbf{Degenerate cases:} Detect degenerate points where trajectories may merge, split, appear, or disappear.
\item \textbf{Particle tracing:} From each boundary point or degenerate case, integrate \cref{eq:ODE} and apply a correction step to generate critical point trajectories. By removing duplicates and connecting discrete trajectories, we incorporate degenerate points and assemble the full set of trajectories.
\end{enumerate}

Any trajectory that does not intersect a degenerate point must intersect the boundary, ensuring full coverage.

The type of a critical point is determined by the spatial Hessian $\mathrm{H}_{\mathbf{x}}f$. In the 2D case:
\begin{enumerate}[noitemsep,leftmargin=*]
    \item Minimum: $\det(\mathrm{H}_{\mathbf{x}}f) > 0,\; f_{x_0 x_0} > 0$.
    \item Maximum: $\det(\mathrm{H}_{\mathbf{x}}f) > 0,\; f_{x_0 x_0} < 0$.
    \item Saddle: $\det(\mathrm{H}_{\mathbf{x}}f) < 0$.
\end{enumerate}
In 3D, critical points are classified based on the eigenvalues of the Hessian matrix, yielding four types: minimum (all eigenvalues positive), maximum (all eigenvalues negative), and two types of saddle (one or two eigenvalues positive).

\subsection{Initialization}
\label{sec:initialization}
We use Newton's method to identify initial critical points on the boundary of the $(d+1)$-dimensional space–time domain, following~\cite{ma2024critical}. The boundary consists of $2d+2$ hyperplanes: $x_0 = x_{0,\min}$, $x_0 = x_{0,\max}$,
$x_1 = x_{1,\min}$, $x_1 = x_{1,\max}$, $\ldots,$ $t = t_{\min}$, $t = t_{\max}$.

Consider the hyperplane $x_0 = x_{0,\min}$. Let $\Tilde{\mathbf{x}}=(x_1,x_2,\cdots)^\top$, we solve for boundary critical points by finding roots of
\begin{equation}
    \mathbf{g}(\Tilde{\mathbf{x}}, t) =
\begin{pmatrix}
f_{x_0}(x_{0,\min}, x_1, x_2, \ldots, t) \\
f_{x_1}(x_{0,\min}, x_1, x_2, \ldots, t) \\
\vdots \\
f_{x_{d-1}}(x_{0,\min}, x_1, x_2, \ldots, t)
\end{pmatrix}=\mathbf{0}.
\end{equation}
The Jacobian of $\mathbf{g}$ is the Hessian of $f$ with respect to the free variables:
\begin{equation}
\label{eq:jacobian}
   J_{\mathbf{g}}(\Tilde{\mathbf{x}},t)
=
\begin{pmatrix}
\displaystyle f_{x_0 x_1} & f_{x_0 x_2} & \cdots & f_{x_0 x_{d-1}} & f_{x_0 t} \\
\displaystyle f_{x_1 x_1} & f_{x_1 x_2} & \cdots & f_{x_1 x_{d-1}} & f_{x_1 t} \\
\vdots & \vdots & \ddots & \vdots & \vdots \\
\displaystyle f_{x_{d-1} x_1} & f_{x_{d-1} x_2} & \cdots & f_{x_{d-1} x_{d-1}} & f_{x_{d-1} t}
\end{pmatrix}.
\end{equation}
A Newton step is given by
\begin{equation}
\label{eq:newton}
    (\Delta \Tilde{\mathbf{x}},\Delta t)
= - J_{\mathbf{g}}^{-1}(\Tilde{\mathbf{x}},t) \, \mathbf{g}(\Tilde{\mathbf{x}},t),
\ \ 
\Tilde{\mathbf{x}} \leftarrow \Tilde{\mathbf{x}} + \Delta \Tilde{\mathbf{x}},\ \ 
t \leftarrow t+\Delta t.
\end{equation}
Since $x_0$ is fixed by the boundary condition, it is excluded from the update. Critical points on other boundaries are updated in the same manner. The iteration terminates once the spatial gradient norm satisfies $||\mathbf{g}(\Tilde{\mathbf{x}}, t)||<\epsilon$.

To initialize Newton’s method, we place candidate starting points along the boundary of the space--time domain. These points are used solely to seed root finding and do not define a discretized scalar field. 

For MFA models, the domain is partitioned into spans; within each boundary span, we uniformly sample $(p+2)^d$ initial points on the boundary hyperplane, where $p$ is the polynomial degree. For analytic functions and INRs, we sample 40 and 80 points, respectively, along the shortest domain dimension and scale the sampling density proportionally along the remaining dimensions, as determined by our ablation study in the supplement. To reduce redundancy, we discard initial points whose spatial distance is below the spatial step size $s_s$ and whose temporal distance is below the time step size $s_t$ simultaneously. The selected sampling densities were further evaluated in the ablation study of the supplement.
As the number of initial points increases, the number of detected seeds converges, suggesting that the chosen densities are sufficient for the datasets considered in this work.
While the convergence behavior provides a practical criterion for assessing seeding adequacy, the current approach does not provide a formal coverage guarantee. Consequently, highly localized or closely spaced structures may require finer or adaptive seeding strategies.

\subsection{Degenerate Cases}
\label{sec:degenerate-case}
Within the space-time domain, critical point trajectories may appear, disappear, split, or merge at degenerate events. We focus on the common case where these events occur as isolated points in space-time, allowing each event to be localized and processed independently. This assumption underlies the hypercube-based tracing procedure described below. More complex configurations involving higher-order degeneracies are outside the scope of the current formulation. 

At degenerate points, the local behavior is non-bijective, implying that the spatial Hessian matrix in \cref{eq:velocity} must be singular. Such points satisfy: 
\begin{equation}
    \mathbf{k}(\mathbf{x},t) =
\begin{pmatrix}
\det(\mathrm{H}_{\mathbf{x}}f) \\
f_{x_0} \\
\vdots \\
f_{x_{d-1}} \\
\end{pmatrix}_{(d+1)\times 1}
= \mathbf{0}.
\end{equation}
We compute degenerate points using Newton’s method:
\begin{equation}
    (\Delta \mathbf{x}, \Delta t)
= - \nabla \mathbf{k}^{-1}(\mathbf{x},t) \, \mathbf{k}(\mathbf{x},t),
\quad
\mathbf{x} \leftarrow \mathbf{x} + \Delta \mathbf{x},
\quad
t \leftarrow t + \Delta t.
\end{equation}
We terminate the iteration when
$
\frac{\lVert \det(H_{\mathbf{x}} f) \rVert}{\lVert H_{\mathbf{x}} f \rVert_F^{\,d}} < \epsilon_k,
$
where degeneracy is evaluated via the scale-invariant normalized determinant, and when
$
\lVert \nabla_{\mathbf{x}} f \rVert < \epsilon.
$

Using Jacobi's formula~\cite{magnus2019matrix}, we express the derivative of the determinant of the Hessian in terms of the adjugate (or adjoint) of the Hessian and its derivative:
\begin{equation}
\frac{\partial}{\partial x_i}\det(\mathrm{H}_{\mathbf{x}}f)
= \operatorname{tr}\!\left(\operatorname{adj}(\mathrm{H}_{\mathbf{x}}f)\,\frac{\partial \mathrm{H}_{\mathbf{x}}f}{\partial x_i}\right)
= \left\langle \operatorname{adj}(\mathrm{H}_{\mathbf{x}}f),\, \frac{\partial \mathrm{H}_{\mathbf{x}}f}{\partial x_i} \right\rangle_F
\end{equation}
where $\operatorname{tr}(\cdot)$ is the trace of the matrix and $\operatorname{adj}(\cdot)$ is its adjugate matrix. As $\mathrm{H}_{\mathbf{x}}f$ is symmetric, the trace is the Frobenius inner product $\langle\cdot, \cdot \rangle_F$.

To perform Newton's method, we sample initial points throughout the entire space–time domain. Following the same principle as in \cref{sec:initialization}, each MFA span is uniformly seeded with $(p+2)^{d+1}$ initial points. For analytic functions and INRs, we again place $40$ and $80$ samples, respectively, along the shortest domain dimension and scale the sampling density proportionally along the remaining dimensions.

\para{Particle tracing from degenerate points.}
To extract trajectories passing through a degenerate point, we construct a small axis-aligned hypercube centered at the point, extending by $s_s$ in each spatial direction and $s_t$ in time (yielding side lengths $2s_s$ and $2s_t$, respectively). On each face of this hypercube, we identify critical points using the procedure described in \cref{sec:initialization}.

For Newton’s method, we uniformly sample $4^d$, $(p+2)^d$, and $8^d$ initial points per face for analytic functions, MFAs, and INRs, respectively. We then remove near-duplicate initializations using a small tolerance, discarding pairs whose spatial distance is below $s_s/10$ and temporal distance below $s_t/10$. These tolerances are chosen heuristically to suppress redundant initializations within the local hypercube.

\begin{figure}[!ht]
\centering
\vspace{-2mm}
\includegraphics[width=\linewidth]{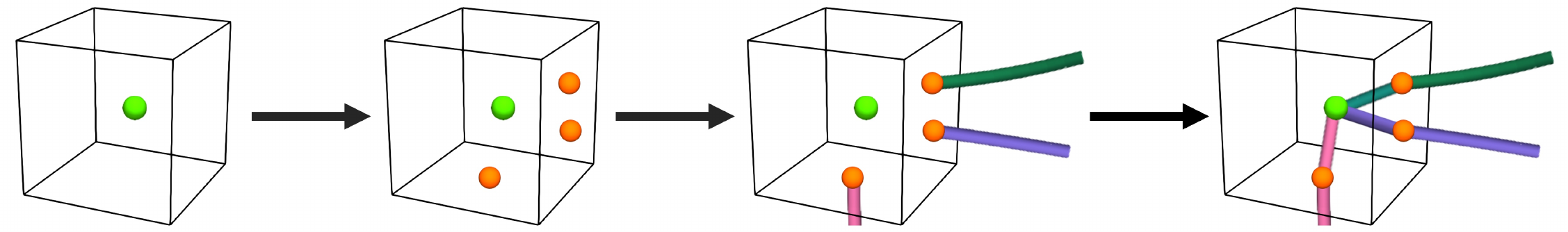}
\vspace{-6mm}
\caption{Degenerate-case handling. The green point marks the degenerate point, and the orange points denote the critical points on the hypercube.}
\label{fig:degenerate-case}
\vspace{-3mm}
\end{figure}

From the resulting critical points, we trace trajectories outward from the hypercube, yielding paths that emanate from the degenerate point. \cref{fig:degenerate-case} illustrates this degenerate-case handling.

\subsection{Particle Tracing}
\label{sec:particle-tracing}

During particle tracing, we balance spatial and temporal integration using prescribed step sizes $s_s$ and $s_t$. Relying solely on the time-parameterized RK4 (\cref{eq:RK4}) provides limited control over spatial progression; therefore, we combine time-parameterized and arc-length--parameterized RK4 schemes (\cref{eq:RK4-spatial}).

\para{RK4 scheme selection.}  
At each location, we select between the two schemes to maintain controlled progression in both space and time. When the velocity magnitude satisfies $|\xi(t,\mathbf{x})| < s_s / s_t$, we first apply a time-parameterized RK4 step and switch to the arc-length formulation if the resulting spatial displacement exceeds $s_s$. Conversely, when $|\xi(t,\mathbf{x})| \ge s_s / s_t$, we begin with the arc-length-parameterized update and revert to the time-parameterized scheme if the temporal increment exceeds $s_t$. This adaptive strategy ensures stable and well-controlled integration.

\para{Tracking direction.}  
Since all starting points lie on boundaries or at degenerate locations, each trajectory is traced in a single direction. For boundary points, we choose the direction that keeps the trajectory within the domain; for degenerate points, we trace away from the degenerate location to ensure proper emergence of trajectories.

\para{Stop conditions.}
As a trajectory approaches a degenerate point, $\det(\mathrm{H}_{\mathbf{x}} f)$ in \cref{eq:velocity} approaches zero, making RK4 updates increasingly unstable and causing both spatial and temporal steps to diminish. We therefore terminate tracing when
\[
|\mathbf{x}_{n+1}-\mathbf{x}_n| < 0.1\, s_s \quad \text{and} \quad |t_{n+1}-t_n| < 0.1\, s_t,
\]
or when the domain boundary is reached. The factor $0.1$ is chosen heuristically to detect when the trajectory has effectively stalled near a degenerate point.
If tracing instead crosses a degenerate point without meeting this stop criterion, we continue tracing and handle the crossing during trajectory splitting.

\para{Correction step.}
Each RK4 step acts as a predictor and can accumulate numerical drift away from the critical-point trajectory.
To reduce integration drift, we refine each predicted point using Newton iterations with $t$ fixed. If $\|\nabla_{\mathbf{x}} f\| > \epsilon$, we iteratively update $\mathbf{x}$ until the critical-point condition is satisfied.

\para{Trajectory splitting.}
If a trajectory crosses a degenerate point in time while remaining within a spatial distance $s_s$, we split the trajectory at that segment into two separate trajectories to correctly capture the degenerate event.

\para{Duplication removal.}
Trajectories are typically seeded from boundaries or degenerate points. Two trajectories $A$ and $B$ are considered duplicates if
\begin{enumerate}[noitemsep,leftmargin=*]
    \item their corresponding endpoints are within $s_s$ spatially and $s_t$ temporally, and a midpoint of one trajectory lies within $(s_s, s_t)$ of the other; or
    \item one endpoint of a trajectory lies within $(s_s, s_t)$ of an endpoint of the other, and its remaining endpoint also lies within $(s_s, s_t)$ of the other trajectory.
\end{enumerate}
If either condition holds, only the longer trajectory is retained as the representative.
Since degenerate points are obtained by numerical root finding from multiple initializations, the same topological event may be detected more than once. We therefore de-duplicate both degenerate points and trajectories using $(s_s, s_t)$. These step sizes define the effective spatial-temporal resolution of the tracking procedure: detections within this resolution are merged, whereas events separated beyond it are retained as distinct.

\para{Connection.}
All detected degenerate points are included in the final output. Trajectory endpoints are connected to a nearby degenerate point when their spatial distance is below $s_s$, and their temporal distance is below $s_t$.

%% file: sec-configuration.tex
\section{Experimental Configuration}
\label{sec:configuration}

We conduct experiments on 2D and 3D time-varying continuous models, including closed-form analytic functions as well as MFA and INR models representing scientific datasets.

\subsection{Evaluation Protocol}

We evaluate critical-point tracking directly on continuous implicit models. Analytic functions have known ground-truth trajectories. For MFA and INR models, where ground truth is unavailable, we assess whether extracted samples satisfy the critical-point condition of the input model. In the supplement, we also compare selected local trajectory structures for the four 2D scientific models with LWM results obtained using denser sampling of the same models.

\para{Baselines.}
For 2D time-varying models, we compare against \emph{Stable Feature Flow Fields} (SFFF)~\cite{weinkauf2011stable} and the \emph{lifted Wasserstein matcher} (LWM)~\cite{soler2018lifted}. To ensure a faithful implementation and fair comparison, we
follow the published SFFF workflow by sampling the continuous implicit model to a gradient vector field. For LWM, we sample each continuous model to a piecewise-linear (PL) scalar field. For 3D time-varying models, we compare against a \emph{discrete extraction} pipeline that samples the model, constructs a PL approximation, and extracts isolated critical points. Applying LWM to the 3D Vortex model would require a $240^3\times160$ grid (approximately 17.7\,GB) before tracking.

\para{Metrics and parameters.}
We report the gradient norm along extracted trajectories and the number of connected components (\#CC). For analytic functions, we additionally compare the number of branches (\#Branch) with ground truth.
We also report the symmetric Hausdorff distance~\cite{huttenlocher1993comparing} between the recovered trajectories and densely sampled ground-truth polylines, as well as the maximum spatial distance between recovered vertices and analytic reference points evaluated at the same times. For MFA and INR models, we evaluate detected degenerate points using the scale-normalized Hessian determinant
$|\det(H_{\mathbf{x}}f)|/\|H_{\mathbf{x}}f\|_F^d$.
We use gradient and Newton thresholds $\epsilon=e^{-10}$ and $\epsilon_k=e^{-6}$, respectively. Spatial and temporal step sizes are selected through convergence studies that progressively decrease the step size until \#CC stabilizes. 
We use $l/k$ for span-relative MFA step sizes and $r/k$ for range-relative analytic and INR step sizes, where $l$ denotes the corresponding span length, $r$ is the shortest domain range, and $k$ is a refinement factor. Larger $k$ gives a smaller step size. Parameterization and sensitivity studies are provided in the supplement.

\para{Complexity and implementation.}
The pipeline runs in
$O((N_{\mathrm{initial}}+N_{\mathrm{start}}/s)E)$,
where $N_{\mathrm{initial}}$ is the number of initial seeds, $N_{\mathrm{start}}$ is the number of tracing start points, and $E$ is the model-evaluation cost. Our C++ implementation uses TBB for analytic and MFA models and LibTorch~\cite{paszke2019pytorch} with autograd and MPI for INR models. We use TTK~\cite{tierny2018topology} for LWM and discrete extraction. Complexity, hardware, and implementation details are given in the supplement.
We refer to our method as \emph{implicit extraction} and the 3D critical point extraction pipeline described above as \emph{discrete extraction}.

\subsection{Continuous Models as Input}

We evaluate our method on both 2D and 3D time-varying continuous models. For 2D, we consider two analytic functions (Quartic Potential and Quartic Rotation) and two scientific datasets (Vortex Street and Heated Cylinder). For each dataset, we construct continuous implicit representations using both MFA and CoordNet models. Although these representations differ in formulation and accuracy, our method operates directly on the continuous models and consistently tracks critical points across them. In contrast to discrete pipelines based on sampled data, our approach is designed for the continuous setting, demonstrating generality across representations.

To assess scalability, we further evaluate 3D time-varying models, including a 3D Quartic Potential function and a 3D Vortex dataset represented using MFA.

\para{2D Quartic Potential} is time-varying over a 2D spatial domain, 
\begin{equation}
    f(x,y,t)=\frac{x^4}{4}+\frac{1-t}{2}x^2+\frac{y^4}{4}+\frac{\cos t}{2}y^2
\end{equation}
defined over $[-2,2]\times[-2,2]\times[0,4]$.
\begin{enumerate}[noitemsep,leftmargin=*]
    \item For $0\le t\le 1$, there is a single critical point at $(0,0,t)$.
    \item For $1<t\le \frac{\pi}{2}$, three critical points appear at $(0,0,t)$ and $(\pm\sqrt{t-1},0,t)$.
    \item For $\frac{\pi}{2}<t\le 4$, nine critical points appear at $(x,y,t)$ with $x\in\{0,\pm\sqrt{t-1}\}$ and $y\in\{0,\pm\sqrt{-\cos t}\}$.
\end{enumerate}
There are four degenerate points: $(0,0,1)$, $(\pm\sqrt{\frac{\pi}{2}-1},0,\frac{\pi}{2})$, and $(0,0,\frac{\pi}{2})$; see \cref{fig:teaser}(A).

\para{3D Quartic Potential} is time-varying over a 3D spatial domain,  
\begin{equation}
    f(x,y,z,t)=\frac{x^4}{4}+\frac{1-t}{2}x^2+\frac{y^4}{4}+\frac{\cos t}{2}y^2+\frac{z^4}{4}+\frac{\cos t}{2}z^2
\end{equation}
over $[-2,2]^3\times[0,4]$.
\begin{enumerate}[noitemsep,leftmargin=*]
    \item For $0\le t\le 1$, there is a single critical point at $(0,0,0,t)$.
    \item For $1<t\le \frac{\pi}{2}$, three critical points appear at $(0,0,0,t)$ and $(\pm\sqrt{t-1},0,0,t)$.
    \item For $\frac{\pi}{2}<t\le 4$, 27 critical points appear at $(x,y,z,t)$ with $x\in\{0,\pm\sqrt{t-1}\}$ and $y,z\in\{0,\pm\sqrt{-\cos t}\}$.
\end{enumerate}
There are four degenerate points: $(0,0,0,1)$, $(\pm\sqrt{\frac{\pi}{2}-1},0,0,\frac{\pi}{2})$, and $(0,0,0,\frac{\pi}{2})$.

\para{2D Quartic Rotation} is a time-varying function on a 2D domain, 
\begin{equation}
\begin{aligned}
    f(x,y,t)=\frac{1}{4}\left(\left(x\cos t+y\sin t\right)^2-1\right)^2 \\+\frac{1}{4}\left(\left(y\cos t-x\sin t\right)^2-1 \right)^2
\end{aligned}
\end{equation}
defined over the domain $[-2,2]\times [-2,2]\times[0,4]$.
It contains nine rotating critical points as illustrated in \cref{fig:teaser}(B), located at 
\begin{equation}
(x,y,t)=(u\cos t-v\sin t,\;u\sin t+v\cos t,\;t),\quad u,v\in\{0,\pm1\}.
\end{equation}

\para{2D Vortex Street.}
We evaluate our method on a simulated von K\'arm\'an vortex street dataset~\cite{popinet2004free,gunther2017generic}, representing 2D viscous flow around a cylinder. We use the velocity magnitude over the final 150 time steps, where the vortex pattern is fully developed. 
For this dataset, the MFA model uses a span configuration of $80\times 10\times 18$, and we additionally train a CoordNet (INR) model with five residual blocks.

\para{2D Heated Cylinder.}
This dataset models 2D flow induced by a heated cylinder under the Boussinesq approximation~\cite{popinet2004free,gunther2017generic}. We use the velocity magnitude over 150 consecutive time steps ($t=1000$ to $t=1149$), excluding the initial transient phase to focus on well-developed flow structures.
The MFA model uses a span configuration of $10\times 30\times 15$, and the CoordNet (INR) model consists of five residual blocks.
Both the Vortex Street and Heated Cylinder datasets include a cylindrical obstacle with zero velocity magnitude. We remove this region in visualizations to reduce clutter and highlight flow structures; results, including the obstacle, are provided in the supplement.

\para{3D Vortex} dataset models vortex structures from a pseudo-spectral simulation~\cite{porter2019a}. We use 90 time steps and select vorticity magnitude as the scalar field. The MFA model uses a span configuration of $15^3 \times 10$.

%% file: sec-analytic-results.tex
\section{Critical Point Tracking from Analytic Functions}
\label{sec:analytic-results}

We first evaluate our method on three analytical time-varying scalar fields with known ground truth. For all three functions, we use a step size of $s = r/40$, as the number of connected components remains stable across varying step sizes; see the supplement for details on parameter selection. 

\cref{tab:evaluation-closed-form} summarizes the results using the selected step size, compared with ground truth. We quantify geometric agreement using the symmetric Hausdorff distance between recovered trajectories and densely sampled ground-truth polylines, as well as the maximum spatial distance between recovered vertices and analytic reference points evaluated at the same times. Details of reference construction and evaluation are provided in the supplement.

\begin{table}[!ht]
\centering
\scriptsize
\setlength{\tabcolsep}{1.45pt}
\renewcommand{\arraystretch}{1.1}
\begin{tabular}{c|cc|cc|c|c|c|c}
\hline
\multirow{2}{*}{Function}
 & \multicolumn{2}{c|}{\#Br.}
 & \multicolumn{2}{c|}{\#CC}
 & \multirow{2}{*}{\makecell{Max Grad.\\Norm}}
 & \multirow{2}{*}{ $H$}
 & \multirow{2}{*}{$E_V$}
 & \multirow{2}{*}{Time(s)} \\
\cline{2-5}
 & GT & Imp
 & GT & Imp
 & & & & \\
\hline
2D Quartic Pot. & 13 & \textbf{13} & 1 & \textbf{1} & $9.52e{-11}$ & $3.48e^{-3}$ & $1.16e^{-4}$ & 0.015 \\
2D Quartic Rot. & 9  & \textbf{9}  & 9 & \textbf{9} & $2.15e{-11}$ & $1.25e^{-3}$ & $2.15e^{-11}$ & 0.021 \\
3D Quartic Pot. & 31 & \textbf{31} & 1 & \textbf{1} & $9.53e{-11}$ & $2.99e^{-3}$ & $9.42e^{-5}$ & 0.113 \\
\hline
\end{tabular}
\vspace{-2mm}
\caption{Evaluation on analytic functions. ``GT'' denotes ground truth and ``Imp'' denotes our implicit extraction method. ``\#Br.'' and ``\#CC'' denote the number of branches and connected components, respectively. $H$ is the symmetric Hausdorff distance (maximum over corresponding components for Quartic Rotation). $E_V$ is the maximum spatial vertex distance at identical times.}
\label{tab:evaluation-closed-form}
\vspace{-3mm}
\end{table}

\begin{figure}[!ht]
\centering
\vspace{-3mm}
\includegraphics[width=\linewidth]{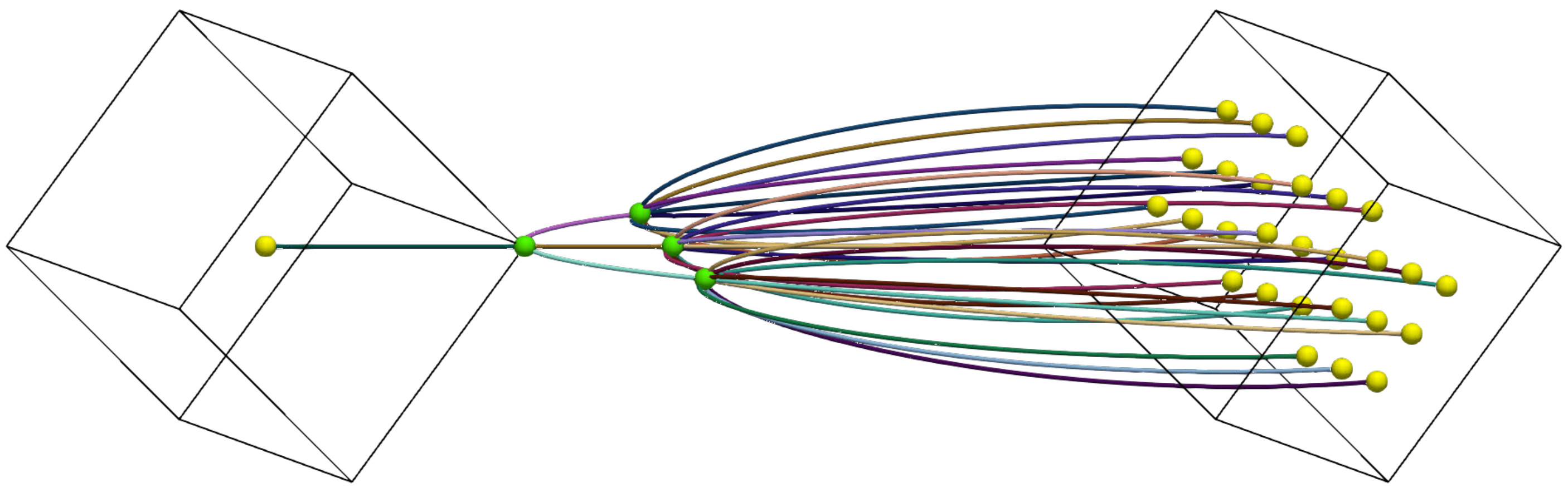}
\vspace{-6mm}
\caption{Tracking results for the 3D Quartic Potential model. Each cube shows the spatial domain at the initial and final time steps. Yellow points denote critical points, and green points indicate degenerate points.}
\label{fig:quartic-potential-3d}
\vspace{-4mm}
\end{figure}

Across all three analytic functions, our method recovers the ground-truth branch and connected-component counts. The small values of $H$ and $E_V$ indicate close geometric agreement with the analytic trajectories.
For the 2D Quartic Potential (\cref{fig:teaser}(A)), trajectory bifurcations and all four degenerate points (in green) are correctly identified. For the 2D Quartic Rotation (\cref{fig:teaser}(B)), we track nine rotating trajectories and illustrate their motion at two representative time steps. For the 3D Quartic Potential (\cref{fig:quartic-potential-3d}), we correctly identify all four degenerate points and recover the corresponding trajectories.

%% file: sec-scientific-results.tex
\section{Critical Point Tracking from MFA and INR Models}
\label{sec:scientific-results}

\begin{table}[t]
\centering
\scriptsize
\renewcommand{\arraystretch}{1.2}
\setlength{\tabcolsep}{1.5pt}

\begin{tabular}{c|c|c|c|cc|cc}
\hline
\multirow{2}{*}{Model} &
\multirow{2}{*}{$s$} &
\multirow{2}{*}{Method} &
\multirow{2}{*}{Data Size} &
\multicolumn{2}{c|}{Grad Norm} &
\multicolumn{2}{c}{$|\det(H)|/\|H\|_F^d$} \\
\cline{5-8}
&&&& Mean & Max & Mean & Max \\
\hline

\multicolumn{8}{c}{\textbf{MFA}}\\

\multirow{3}{*}{\makecell[c]{2D\\Vortex\\Street}}
& \multirow{3}{*}{$\frac{l}{16}$}
& Ours
& \textbf{356 KB}
& \scalebox{0.9}[1]{\boldmath $2.16e^{-12}$}
& \scalebox{0.9}[1]{\boldmath $9.98e^{-11}$}
& \scalebox{0.9}[1]{\boldmath $2.80e^{-11}$}
& \scalebox{0.9}[1]{\boldmath $1.36e^{-9}$}\\

&& SFFF
& 1.41 GB
& $1.16e^{-2}$
& 0.545
& $6.86e^{-2}$
& 0.497\\

&& LWM
& 430 MB
& $7.92e^{-2}$
& 0.998
& 0.215
& 0.500\\
\hline

\multirow{3}{*}{\makecell[c]{2D\\Heated\\Cylinder}}
& \multirow{3}{*}{$\frac{l}{16}$}
& Ours
& \textbf{91.3 KB}
& \scalebox{0.9}[1]{\boldmath $1.30e^{-11}$}
& \scalebox{0.9}[1]{\boldmath $1.00e^{-10}$}
& \scalebox{0.9}[1]{\boldmath $1.94e^{-11}$}
& \scalebox{0.9}[1]{\boldmath $3.51e^{-9}$}\\

&& SFFF
& 295 MB
& $6.35e^{-3}$
& 0.477
& $7.01e^{-2}$
& 0.497\\

&& LWM
& 92.1 MB
& $4.78e^{-2}$
& 0.826
& 0.193
& 0.500\\
\hline

\multirow{2}{*}{\makecell[c]{3D\\Vortex}}
& \multirow{2}{*}{$\frac{l}{16}$}
& Ours
& \textbf{1.21 MB}
& \scalebox{0.9}[1]{\boldmath $3.70e^{-12}$}
& \scalebox{0.9}[1]{\boldmath $1.00e^{-10}$}
& \scalebox{0.9}[1]{\boldmath $5.33e^{-13}$}
& \scalebox{0.9}[1]{\boldmath $1.69e^{-10}$}\\
&& Discrete
& 17.7 GB
& $4.78e^{-3}$
& 0.113
& --
& --\\
\hline

\multicolumn{8}{c}{\textbf{INR}}\\
\multirow{3}{*}{\makecell[c]{2D\\Vortex\\Street}}
& \multirow{3}{*}{$\frac{r}{80}$}
& Ours
& \textbf{6.64 MB}
& \scalebox{0.9}[1]{\boldmath $1.38e^{-12}$}
& \scalebox{0.9}[1]{\boldmath $9.99e^{-11}$}
& \scalebox{0.9}[1]{\boldmath $5.23e^{-10}$}
& \scalebox{0.9}[1]{\boldmath $2.72e^{-8}$}\\

&& SFFF
& 147 MB
& 0.412
& 113
& 0.169
& 0.500\\

&& LWM
& 46.2 MB
& 0.872
& 88.9
& 0.164
& 0.500\\
\hline

\multirow{3}{*}{\makecell[c]{2D\\Heated\\Cylinder}}
& \multirow{3}{*}{$\frac{r}{80}$}
& Ours
& \textbf{6.64 MB}
& \scalebox{0.9}[1]{\boldmath $1.23e^{-11}$}
& \scalebox{0.9}[1]{\boldmath $9.99e^{-11}$}
& \scalebox{0.9}[1]{\boldmath $1.32e^{-10}$}
& \scalebox{0.9}[1]{\boldmath $1.97e^{-8}$}\\

&& SFFF
& 55.2 MB
& 2.14
& 127
& 0.121
& 0.500\\

&& LWM
& 17 MB
& 5.02
& 141
& 0.169
& 0.500\\
\hline
\end{tabular}
\vspace{-2mm}
\caption{Evaluation on MFA and INR models. We report the gradient norm at all trajectory points and the scale-normalized Hessian determinant, $|\det(H)|/\|H\|_F^d$, at all detected degenerate points. The step size is denoted by $s$.}
\vspace{-4mm}
\label{tab:mfa-inr-results}
\end{table}

We now evaluate our method on three MFA models and two INR models derived from scientific datasets.
The supplement provides additional comparisons with LWM in densely resampled local regions.

As summarized in \cref{tab:mfa-inr-results}, the maximum spatial gradient norm of implicit extraction remains below $\epsilon$ for every MFA and INR model. Because the maximum is computed over all extracted trajectory points, this establishes that every reported implicit sample satisfies the defining critical-point condition of the input continuous model within the prescribed tolerance. SFFF and LWM operate on sampled representations, and their extracted points generally have larger residuals when evaluated using the original continuous model. This comparison measures adherence to the critical-point condition of the continuous model rather than complete trajectory correctness.
Implicit extraction also requires substantially less memory because it does not construct a densely sampled surrogate.
For the scale-normalized Hessian determinant, smaller values indicate closer adherence to the Hessian-singularity condition used to define degenerate events. Implicit extraction produces smaller mean and maximum values than SFFF and LWM in the reported experiments. SFFF does not explicitly compute degenerate points. Instead, these points are inferred from the trajectory tracing process. LWM also extracts degenerate points from discrete sampling.

\begin{figure}[!ht]
\centering
\includegraphics[width=1.0\linewidth]{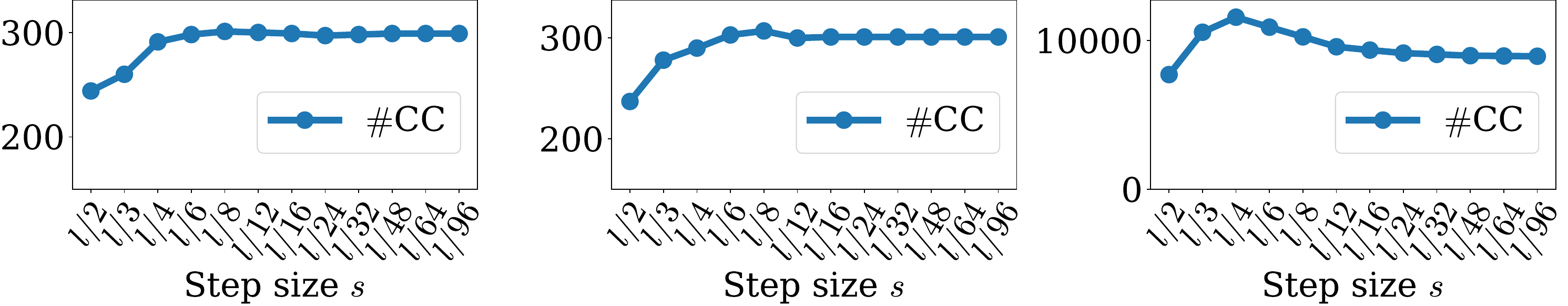}
\vspace{-6mm}
\caption{\#CC versus step size $s$ for MFA models: 2D Vortex Street (left), 2D Heated Cylinder (center), and 3D Vortex (right).}
\label{fig:mfa_step_size_selection}
\vspace{-4mm}
\end{figure}

We select the step size based on the stability of the number of connected components (\#CC), as shown in \cref{fig:mfa_step_size_selection}. Accordingly, we use $s = l/16$ for all three MFA models. A similar strategy is applied to the two INR models, where we use $s = r/80$ (see supplement for details).

\begin{figure*}[!ht]
\centering
\vspace{-3mm}
\includegraphics[width=1.0\linewidth]{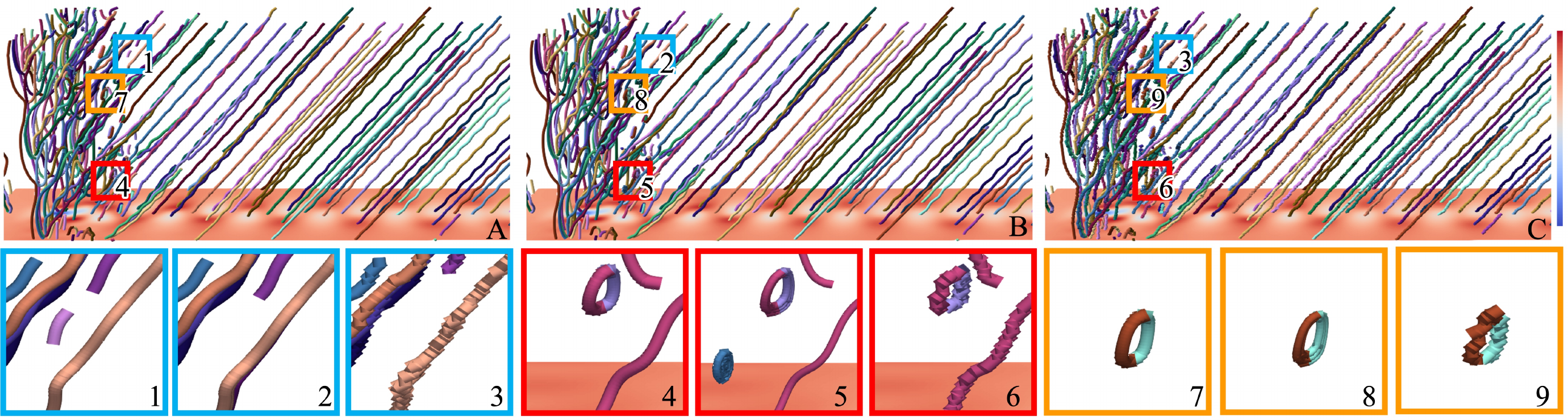}
\vspace{-6mm}
\caption{Vortex Street MFA model: Critical-point trajectories extracted by implicit extraction (A), SFFF (B), and LWM (C). The bottom row presents the zoom-in view of the regions highlighted by the colored boxes in the corresponding top figures. Blocks (1, 4, 7) correspond to (A), (2, 5, 8) to (B), and (3, 6, 9) to (C).}
\label{fig:vortex-street-mfa}
\vspace{-2mm}
\end{figure*}

\begin{figure*}[!ht]
\centering
\vspace{-2mm}
\includegraphics[width=1\linewidth]{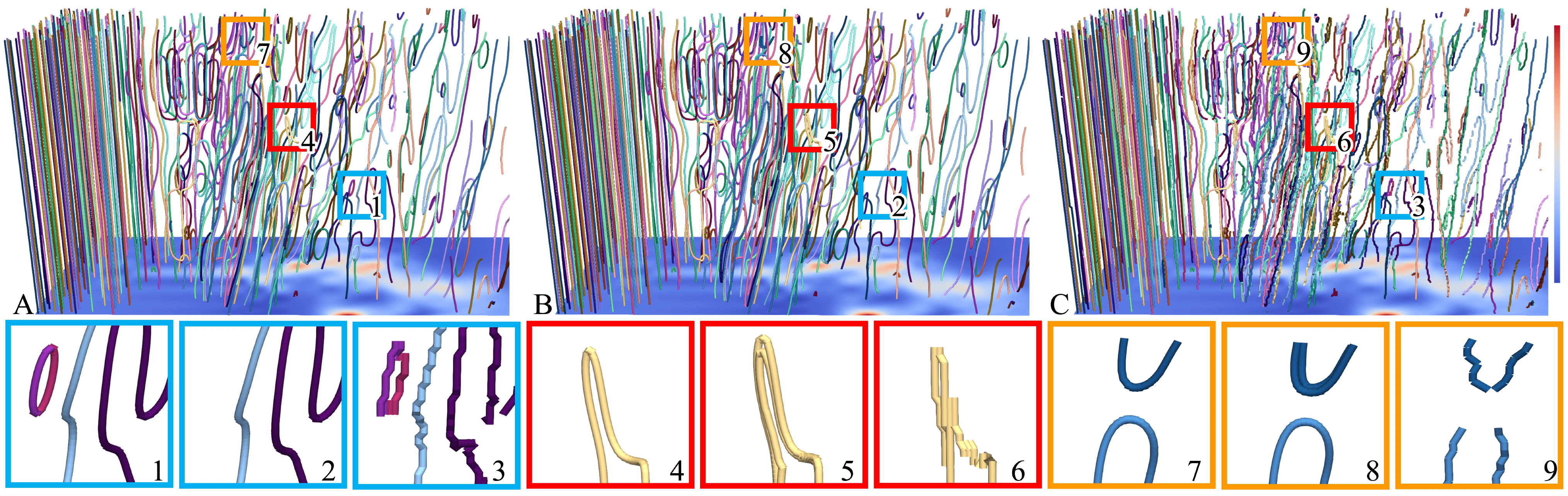}
\vspace{-6mm}
\caption{Heated Cylinder MFA model: Critical-point trajectories extracted by implicit extraction (A), SFFF (B), and LWM (C). The bottom row presents the zoom-in view of the regions highlighted by the colored boxes in the corresponding top figures. Blocks (1, 4, 7) correspond to (A), (2, 5, 8) to (B), and (3, 6, 9) to (C).}
\label{fig:heated-cylinder-mfa}
\vspace{-5mm}
\end{figure*}

\subsection{Tracking Results for 2D MFA Models}

For the two 2D MFA models, \cref{tab:mfa-inr-results} shows that implicit extraction maintains spatial gradient norms below $\epsilon$ and substantially reduces the required data size relative to other two methods.

\para{Vortex Street MFA model.}
In the Vortex Street MFA model (\cref{fig:vortex-street-mfa}), implicit extraction and SFFF produce broadly similar trajectory structures at the global scale. The enlarged views reveal several local differences. LWM produces visibly zigzag trajectories in the highlighted regions. In zoom-in blocks (5) and (8), SFFF reconstructs a trajectory with multiple spiral turns, whereas implicit extraction recovers a closed loop in blocks (4) and (7). All points from the implicit extraction, including those in the highlighted regions, satisfy $\|\nabla_{\mathbf{x}}f\|<\epsilon$.

\para{Heated Cylinder MFA model.}
In the Heated Cylinder MFA model (\cref{fig:heated-cylinder-mfa}), implicit extraction and SFFF again produce broadly similar trajectory structures. The enlarged views reveal several local differences. In blocks (1) and (2), implicit extraction recovers a small closed loop that is not recovered by SFFF at the tested sampling resolution. Every sample on this loop satisfies the critical-point condition of the continuous model within $\epsilon$. In blocks (4) and (5), and again in blocks (7) and (8), implicit extraction produces one continuous trajectory, whereas SFFF produces two closely spaced trajectories.

\subsection{Tracking Results from 2D INR Models}
We apply implicit extraction to two INR models. As in \cref{tab:mfa-inr-results}, the spatial gradient norm along the extracted trajectories remains below $\epsilon$.

\begin{figure*}[!ht]
\centering
\vspace{-2mm}
\includegraphics[width=1.0\linewidth]{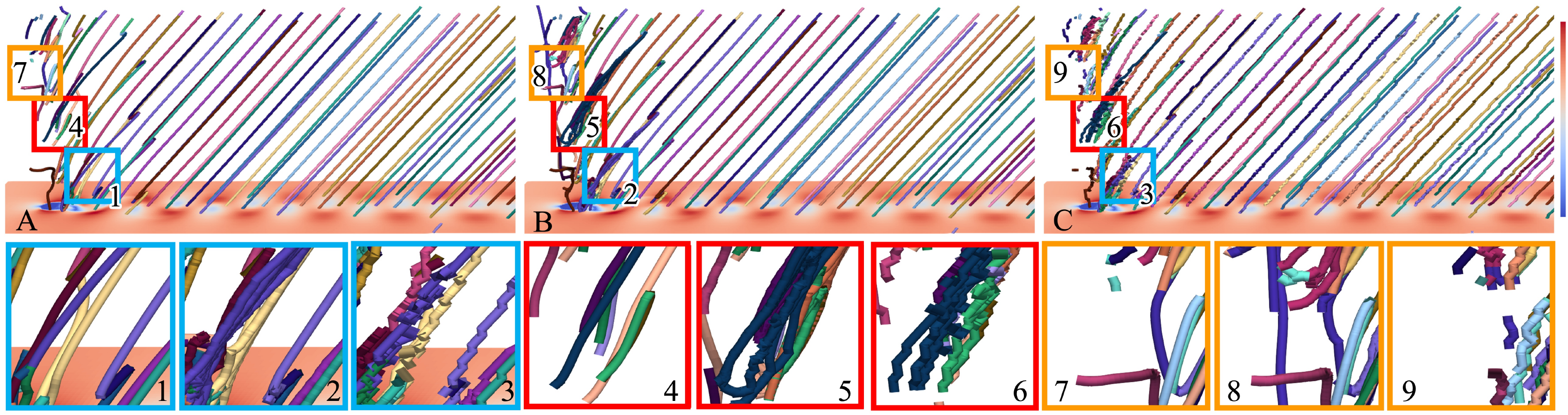}
\vspace{-6mm}
\caption{Vortex Street INR model: Critical-point trajectories extracted by implicit extraction (A), SFFF (B), and LWM (C). The bottom row presents the zoom-in view of the regions highlighted by the colored boxes in the corresponding top figures. Blocks (1, 4, 7) correspond to (A), (2, 5, 8) to (B), and (3, 6, 9) to (C).}
\label{fig:vortex-street-inr}
\vspace{-2mm}
\end{figure*}

\para{Vortex Street INR model.}
As shown in \cref{fig:vortex-street-inr}, the three methods produce broadly similar trajectory structures at the global scale, while the zoom-in views reveal local differences. In the first blue set of blocks (1-3), implicit extraction reconstructs one smooth deep-purple trajectory, SFFF produces multiple closely spaced and nearly parallel trajectories, and LWM produces a visibly zigzag trajectory. Similar differences appear in the second red set of blocks (4-6). In the final orange set of blocks (7-9), SFFF produces an additional purple trajectory that is absent from both implicit extraction and LWM. All implicit extraction trajectory samples satisfy $\|\nabla_{\mathbf{x}}f\|<\epsilon$.

In general, INR models incur higher computational cost than MFA models due to more expensive derivative evaluations. 
While MFA representations provide efficient access to function values and derivatives, INR-based tracking requires repeated neural-network evaluations and derivative computations, resulting in longer time as reflected in \cref{tab:time}.

\begin{table}[!ht]
\centering
\scriptsize
\vspace{-2mm}
\setlength{\tabcolsep}{1.5pt}
\begin{tabular}{cc|cc}
\hline
Model & Time (s) & Model & Time (s) \\
\hline
2D Vortex Street MFA      & 15.3  & 2D Vortex Street INR      & 585  \\
2D Heated Cylinder MFA    & 8.47  & 2D Heated Cylinder INR    & 2151 \\
3D Vortex MFA             & 5938  &                           &         \\
\hline
\end{tabular}
\vspace{-2mm}
\caption{Running time on different models.}
\vspace{-3mm}
\label{tab:time}
\end{table}

A key advantage of the implicit method is that it avoids constructing a dense sampled surrogate, as required by discrete extraction. As the step size decreases, the cost of trajectory tracing grows approximately linearly with the number of sampled points, whereas discrete sampling of the full domain scales cubically in a 2D time-varying setting. In our experiments, running implicit extraction at $s = r/960$ would require an equivalent discrete method to use a grid of size $7680 \times 960 \times 1440$, amounting to more than $84.9$ GB of data. In practice, this discrete approach would necessitate processing each time slice independently and extracting critical points slice by slice.

\begin{figure*}[!ht]
\centering
\vspace{-2mm}
\includegraphics[width=\linewidth]{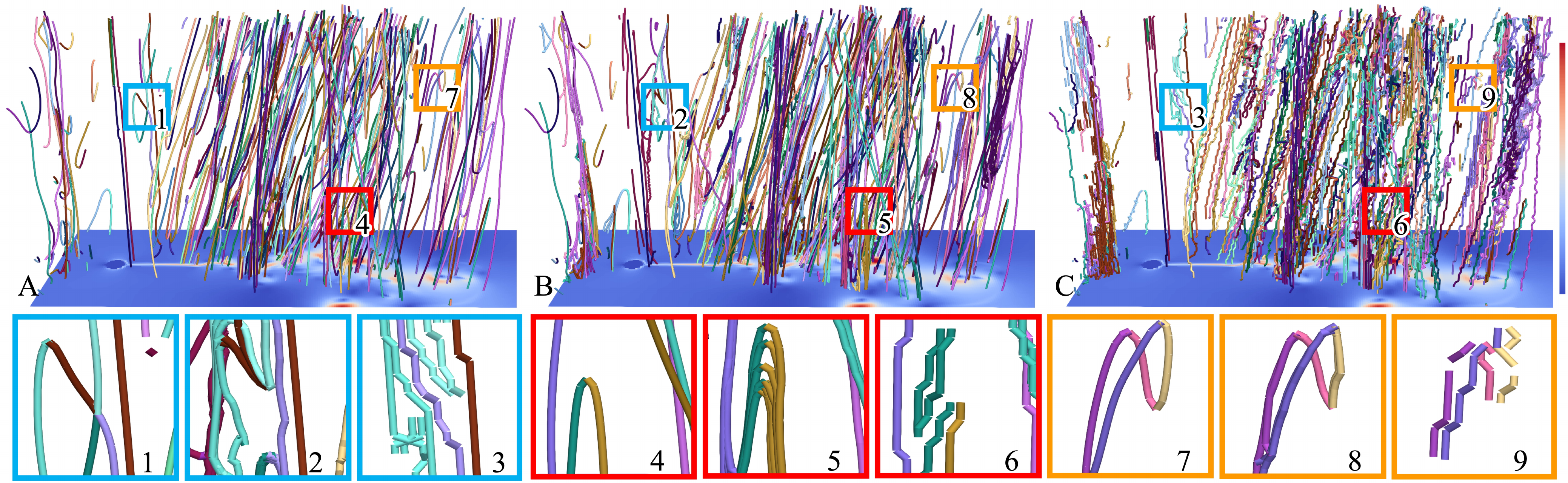}
\vspace{-6mm}
\caption{Heated Cylinder INR model: Critical-point trajectories extracted by implicit extraction (A), SFFF (B), and LWM (C). The bottom row presents the zoom-in view of the regions highlighted by the colored boxes in the corresponding top figures. Blocks (1, 4, 7) correspond to (A), (2, 5, 8) to (B), and (3, 6, 9) to (C).}
\label{fig:heated-cylinder-inr}
\vspace{-5mm}
\end{figure*}

\para{Heated Cylinder INR model.}
A similar comparison is shown for the Heated Cylinder INR model in \cref{fig:heated-cylinder-inr}. In the first blue set of blocks (1-3), implicit extraction identifies a degenerate point at which four trajectories meet, while the corresponding SFFF result does not represent this four-way junction and LWM produces visibly zigzag trajectories. The normalized Hessian determinant reported in \cref{tab:mfa-inr-results} confirms that the degenerate points identified by implicit extraction satisfy the adopted degeneracy criterion within the prescribed tolerance. In the second red set of blocks (4-6), implicit extraction produces smooth trajectories, whereas SFFF produces partially overlapping nearby trajectories and LWM again produces jagged trajectories. A similar pattern appears in the lower-left portion of the final orange set of blocks (7-9).

\subsection{Tracking Results for 3D MFA Model}

\begin{figure*}[!ht]
\centering
\includegraphics[width=0.7\linewidth]{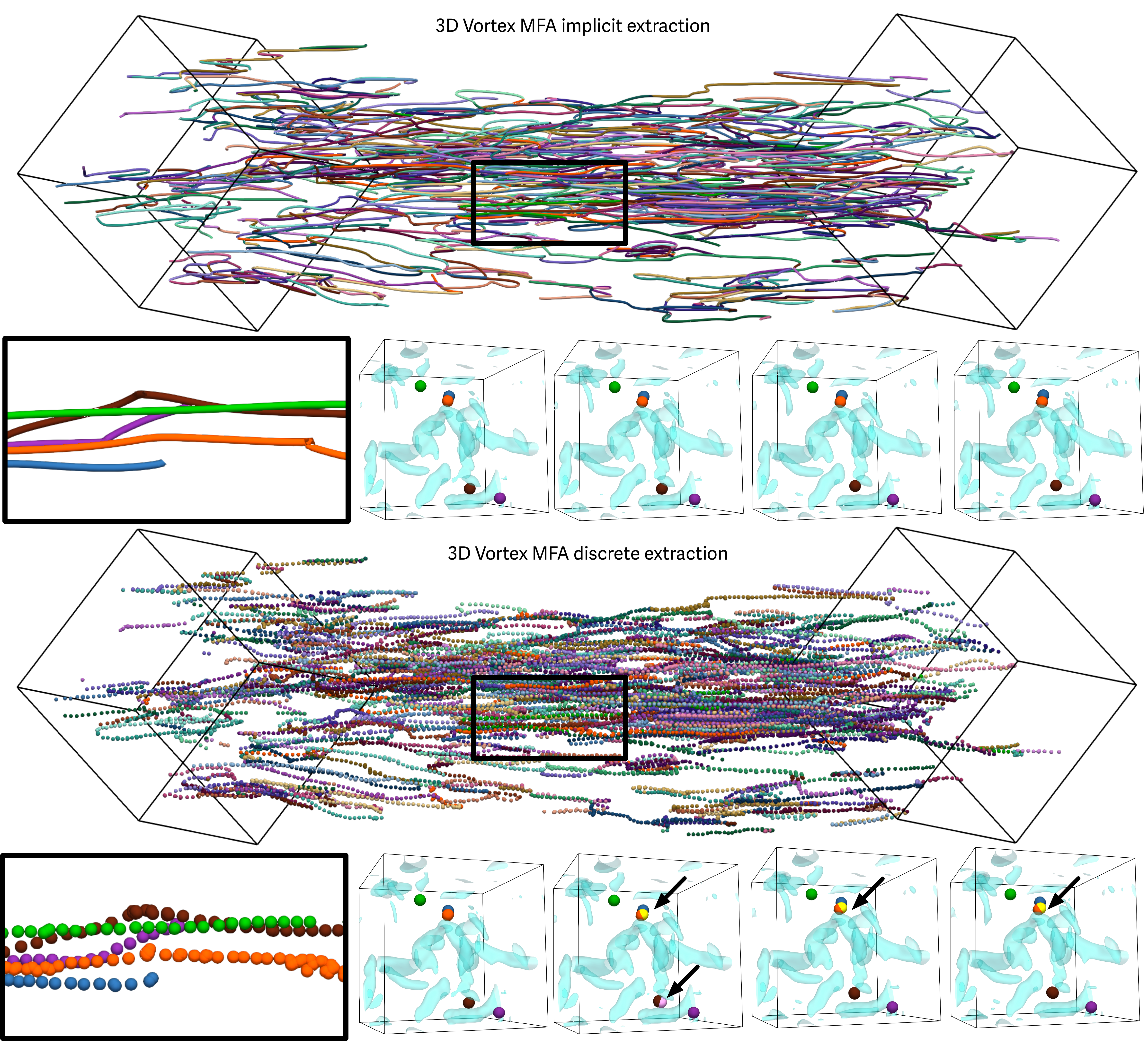}
\vspace{-4mm}
\caption{3D Vortex MFA model: Critical-point trajectories obtained via implicit extraction (top) and discrete extraction (bottom). Insets show magnified views of selected trajectories, along with the corresponding critical-point locations over four consecutive time steps.}
\label{fig:vortex-mfa}
\vspace{-7mm}
\end{figure*}

We demonstrate that our framework generalizes to 3D time-varying data using the 3D Vortex MFA model. 

As shown in \cref{fig:vortex-mfa}, implicit and discrete extraction recover comparable large-scale critical-point structures. The magnified views reveal local differences: at the selected time steps, discrete extraction produces multiple closely spaced critical-point detections, indicated by black arrows, whereas implicit extraction produces one corresponding point. These results are obtained from different computational representations: implicit extraction directly queries the continuous MFA model, whereas discrete extraction operates on a sampled PL approximation. All implicit points satisfy $\|\nabla_{\mathbf{x}}f\|<\epsilon$. As reported in \cref{tab:mfa-inr-results}, implicit extraction also avoids constructing the 17.7~GB sampled representation required by the discrete pipeline.

This model is the most computationally demanding case in this study. 
Its higher-dimensional space requires substantially more boundary initialization, degenerate-point detection, and trajectory tracing than the 2D cases, which accounts for the longer runtime in \cref{tab:time}. Our goal is to demonstrate feasibility on challenging higher-dimensional data, not to claim real-time performance optimization.

%% file: sec-conclusion.tex
\section{Conclusion}
\label{sec:conclusion}

We present a framework for tracking critical points directly on continuous implicit models with access to function values and derivatives. 
By querying the continuous model directly, our approach avoids additional grid resampling while preserving the spatial and temporal coherence of features. Experiments on analytic functions, MFA models, and INR representations across both synthetic and scientific datasets demonstrate robust and accurate tracking of critical-point trajectories, including bifurcations.
More broadly, this work establishes a foundation for topological data analysis on continuous implicit models and opens the door to integrating topology-aware operations into model representations and learning pipelines, enabling feature-aware optimization, topology-consistent learning, and real-time feature tracking. Our approach provides an alternative when a continuous model is already available and dense resampling is costly.

\para{Limitations.} Evaluating derivatives of INR models remains computationally expensive because it requires repeated backpropagation. Our method may be less reliable in regions where the field is constant or nearly constant and critical structure is ill-defined. Parameter selection relies on a convergence study and may require multiple runs for previously unseen datasets. Trajectory extraction also depends on seed placement and may miss features if coverage is insufficient, motivating principled or adaptive seeding strategies.
More generally, our framework does not guarantee complete trajectory recovery:
although we query the continuous model directly, we represent trajectories as a discrete graph. Stable connected-component counts provide a heuristic for selecting the step size but do not establish convergence of the full graph. Because connection and deduplication tolerances depend on the step size, refinement may alter connectivity within existing components. Branches smaller than the step size may also remain unresolved.

%% file: sec-ack.tex
\acknowledgments{
This material is based upon work supported in part by the U.S. Department of Energy (DOE), Office of Science, Office of Advanced Scientific Computing Research, under Contract DE-AC02-06CH11357 and Grants DE-SC0023157 and DE-SC0023145, and by the National Science Foundation (NSF) under Grant DMS-2301361. We used \textsf{ChatGPT 5.6} solely for grammar correction and language polishing.
}

%% file: sec-app-step-size-hardware.tex
This supplement provides details on step-size parameterization (\cref{sec:step-size-details}), hardware and implementation (\cref{sec:software-hardware}), the central finite-difference formulation (\cref{sec:finite-difference-approximation}), computational complexity (\cref{sec:computational-complexity}), and parameter selection for both analytic functions (\cref{sec:analytic-parameters}) and scientific datasets (\cref{sec:scientific-parameters}).
It also examines the necessity of Newton correction (\cref{sec:necessity-of-newton-correction}) and explicit handling of degenerate cases (\cref{sec:necessity-of-degenerate-cases}). It further validates the method against analytic trajectory references (\cref{sec:analytic-reference-validation}) and through local reference comparisons for scientific models (\cref{sec:INR-MFA-validation}).
Additional experimental results are provided in \cref{sec:additional-results}.

\section{Step-Size Parameterization}
\label{sec:step-size-details}
We select spatial and temporal step sizes through convergence studies,
progressively refining them until the number of connected components
(\#CC) stabilizes. For MFA models, the candidate step sizes are
\[
s_s=l_s/k,\qquad s_t=l_t/k,\qquad
k\in\{2,3,4,6,8,\ldots\},
\]
where $l_s$ and $l_t$ are the corresponding spatial and temporal span lengths.

For analytic functions and INR models, we use
\[
s_s=s_t=r/k,\qquad
k\in\{20,30,40,60,80,\ldots\},
\]
where $r$ is the shortest domain range. Equal spatial and temporal step sizes are used only when their
scales are comparable. The method does not otherwise require
$s_s=s_t$. Detailed convergence and sensitivity results are provided in
\cref{sec:analytic-parameters,sec:scientific-parameters}.

\section{Hardware and Implementation}
\label{sec:software-hardware}
The implementation is compiled with \texttt{g++} 11.4.0 using
\texttt{-O3}. All performance measurements are conducted on a desktop
with an eight-core Intel i9 CPU at 3.5\,GHz and 32\,GB of DDR4 memory.

Analytic functions and MFA models are parallelized using Intel Threading
Building Blocks (TBB), with one thread per physical core. INR models are
evaluated using LibTorch, the C++ backend of
PyTorch, with MPI used for multiprocessing.
First-order INR derivatives are computed using
\texttt{torch::autograd::grad()}, while higher-order derivatives are
approximated by central finite differences; see
\cref{sec:finite-difference-approximation}.

We use the Topology ToolKit (TTK) as the baseline implementation for methods that operate on piecewise-linear (PL) scalar fields. 
For LWM, we apply \texttt{TTKTrackingFromFields()}.
For 3D time-varying experiments, we triangulate each 3D volume independently and extract critical points using \texttt{ttkScalarFieldCriticalPoints()}. 
For SFFF, we implemented the algorithm in C++ following the pipeline described in the original paper.

%% file: sec-app-finite-difference.tex
\section{Central Finite-Difference Formula}
\label{sec:finite-difference-approximation}

Given a scalar function $f$, if $g_i$ is the $i$-th component of the gradient of $f$, we approximate second- and third-order derivatives as:  
\begin{equation}
\frac{\partial^2 f}{\partial x_i \partial x_j} \approx
\frac{
g_i(\mathbf{x} + h\mathbf{e}_j)
- g_i(\mathbf{x} - h\mathbf{e}_j)
}{2h},
\end{equation}
\begin{equation}
\begin{aligned}
    \frac{\partial^3 f}{\partial x_i \partial x_j \partial x_k} \approx &
\frac{
\frac{\partial^2 f}{\partial x_i \partial x_j}(\mathbf{x} + h\mathbf{e}_k)
-\frac{\partial^2 f}{\partial x_i \partial x_j}(\mathbf{x} - h\mathbf{e}_k)
}{2h} \\
=&\frac{
g_i(\mathbf{x} + h\,\mathbf{e}_j + h\,\mathbf{e}_k)
- g_i(\mathbf{x} - h\,\mathbf{e}_j + h\,\mathbf{e}_k)
}{4h^{2}} \\
&-\frac{
g_i(\mathbf{x} + h\,\mathbf{e}_j - h\,\mathbf{e}_k)
- g_i(\mathbf{x} - h\,\mathbf{e}_j - h\,\mathbf{e}_k)
}{4h^{2}}.
\end{aligned}
\end{equation}

Here, $h$ is a small perturbation step, and $\mathbf{e}_i$, $\mathbf{e}_j$, and $\mathbf{e}_k$ denote the canonical unit vectors along their respective coordinate directions. Both approximations incur a truncation error of $O(h^2)$. In our experiments, the input coordinates of CoordNet models are normalized to the domain $[-1,1]^3$, and we set $h = e^{-4}$ for finite-difference estimation of higher-order derivatives.

We deliberately combine automatic differentiation for first-order derivatives with central finite differences for higher-order derivatives. Accurate first-order gradients from \texttt{autograd} are critical for keeping trajectories aligned with true critical points, whereas enabling higher-order \texttt{autograd} would introduce significant memory and computational overhead. Central finite differences offer a simple and efficient alternative. In practice, this hybrid strategy yields stable higher-order derivative estimates that are sufficient for reliably tracking critical points.

To analyze the stability of this approach, let $\hat{H} = H + E$ denote the finite-difference approximation of the Hessian. Since the central-difference scheme is second-order accurate, the perturbation satisfies $E = O(h^2)$ entry-wise.

Using Jacobi's formula, the differential of a determinant is:
\begin{equation}
  d(\det(H)) = \text{tr}(\text{adj}(H) \, dH),  
\end{equation}
where $\text{adj}(H)$ is the adjugate matrix of $H$.
It follows that
\begin{equation}
    \begin{aligned}
        |\det(\hat H)-\det(H)| &= |\text{tr}(\text{adj}(H) E) + O(\|E\|_F^2) | \\ 
        &\leq \|\text{adj}(H)\|_F \|E\|_F + O(\|E\|_F^2)\\
        &=  O\!\left(\|H\|_F^{\,d-1} h^2\right),
    \end{aligned}
\end{equation}
where $d$ is the spatial dimension and $\|\cdot\|_F$ denotes the Frobenius norm.
Consequently, dividing by $\|H_{\mathbf{x}}f\|_F^{\,d}$ safely mitigates the $O(\|H\|_F^{d-1})$ scaling of the absolute error, ensuring that the scale-normalized determinant $\frac{|\det(H_{\mathbf{x}}f)|}{\|H_{\mathbf{x}}f\|_F^{\,d}}$ remains stable up to the truncation error induced
by finite differencing. 

In our experiments, we use the same scale-invariant normalized determinant criterion for both exact and finite-difference Hessians. The error analysis above provides a principled theoretical basis for selecting the threshold $\epsilon_k$ in the approximate setting. Empirically, our ablation study in \cref{fig:degenerate_points_with_epsilon_k} shows that the number of detected degenerate points remains unchanged across a range of $\epsilon_k$ values, indicating that the method is robust to the precise choice of threshold. 
Accordingly, we terminate Newton iteration for degenerate point detection when
\[
\frac{|\det(H_{\mathbf{x}}f)|}{\|H_{\mathbf{x}}f\|_F^{\,d}} < \epsilon_k
\quad\text{and}\quad
\|\nabla_{\mathbf{x}}f\| < \epsilon.
\]

%% file: sec-app-complexity.tex
\section{Computational Complexity}
\label{sec:computational-complexity}

We first generate $N_{\mathrm{initial}} = \prod_{i=1}^{d+1} N_i$ initial seeds to locate degenerate points using Newton's method, where $d+1$ denotes the spatial dimension plus time. This stage requires $O(N_{\mathrm{initial}})$ evaluations of the model and its derivatives. For analytical functions and INR models, we set $\min\{N_i\}=40, 80$, respectively, and scale the other dimensions proportionally. For MFA models, we choose $N_i=(p+2)N_{\mathrm{span},i}$, where $p$ is the spline degree ($p=3$ in all our experiments) and $N_{\mathrm{span},i}$ is the number of spans in the $i$-th dimension.

To obtain start points for particle tracing, we sample along the domain boundaries, requiring $O\left(\sum_{i=1}^{d+1} \prod_{j \neq i} N_j\right)$ model evaluations. Suppose we obtain $D$ degenerate points ($D \ll N_{\mathrm{initial}}$) after degenerate point finding. Around each degenerate point, we sample initial points from a small hypercube to find additional start points. For analytical functions and INRs, we place $4^d$ and $8^d$ initial points, respectively, on each face of the hypercube, and for MFA, we use $(p+2)^d$ initial points per face to initialize Newton’s method. This adds $O(D)$ extra evaluations. 

Overall, detecting all start points costs $O(N_{\mathrm{initial}})$ evaluations of the model and its derivatives.

Let $N_\mathrm{start}$ denote the number of start points. Each trajectory is integrated using RK4 with step size $s$, requiring $O(1/s)$ steps. Thus, the total number of traced samples is $O(N_{\mathrm{start}}/s)$. Splitting, deduplication, and connection use spatial hashing, yielding expected linear time in the number of traced samples.

Therefore, the total computational complexity of our pipeline is $O((N_{\mathrm{initial}}+N_{\mathrm{start}}/s)E)$ where $E$ is the cost of a single evaluation of value and gradient. For MFA, $E=O(p^d)$, which is effectively constant for a fixed degree and a fixed dimension. For INR, $E=O(LW^2)$ for a network with depth $L$ and width $W$, dominated by dense matrix multiplications during forward–backward evaluation.

%% file: sec-app-analytic-parameters.tex
\section{Parameter Justification for Analytic Functions}
\label{sec:analytic-parameters}

\begin{figure}[!ht]
\centering
\includegraphics[width=1.0\linewidth]{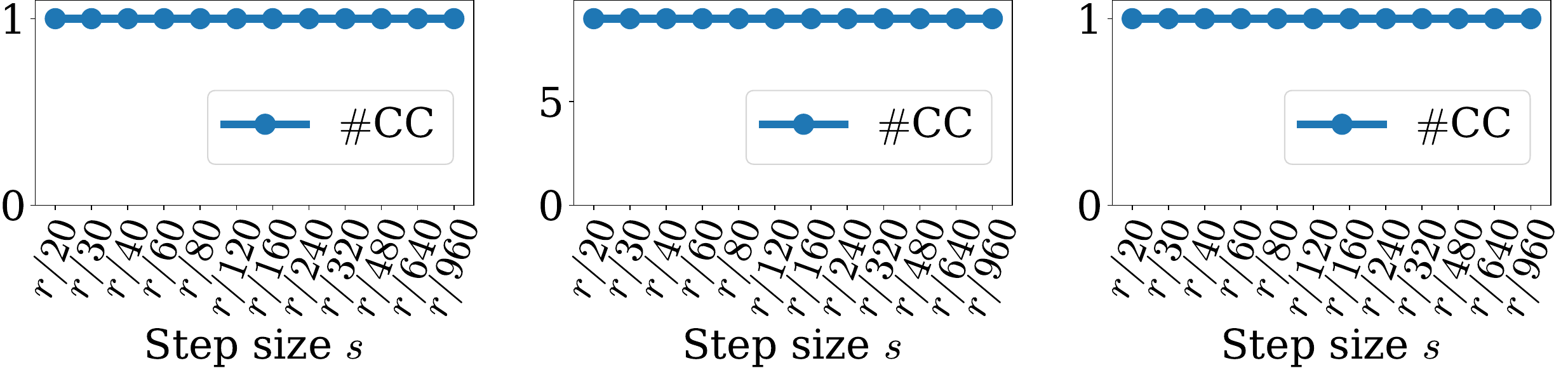}
\vspace{-6mm}
\caption{\#CC vs. step size $s$ for 2D Quartic Potential (left), 2D Quartic Rotation (center), and 3D Quartic Potential (right).}
\label{fig:closd_form_step_size_selection}
\vspace{-3mm}
\end{figure}

For all three analytic functions, we choose the same step size $s = r/40$ as the number of connected components remains stable across different step sizes,
as shown in \cref{fig:closd_form_step_size_selection}. 

%% file: sec-app-scientific-parameters.tex
\section{Parameter Justification for Scientific Datasets}
\label{sec:scientific-parameters}

This section provides empirical justification for the parameter settings used in our scientific datasets, including the step size and ablation studies on the number of seeds, $\epsilon$, and $\epsilon_k$. We evaluate these choices using several metrics: the spatial gradient norm (Grad Norm), which measures how well the extracted trajectories satisfy the critical-point condition; and the number of loops and connected components (\#Loop and \#CC), which reflect the topological stability of the extracted trajectories. 

For the Vortex Street and Heated Cylinder MFA and INR models, we report results for both obstacle-free and obstacle-inclusive settings when applicable. Since the meaningful domain is the obstacle-free region, our analysis primarily focuses on the obstacle-free setting.

\subsection{Parameter Selection of Step Size for INR Models}

\begin{figure}[!ht]
\centering
\includegraphics[width=1.0\linewidth]{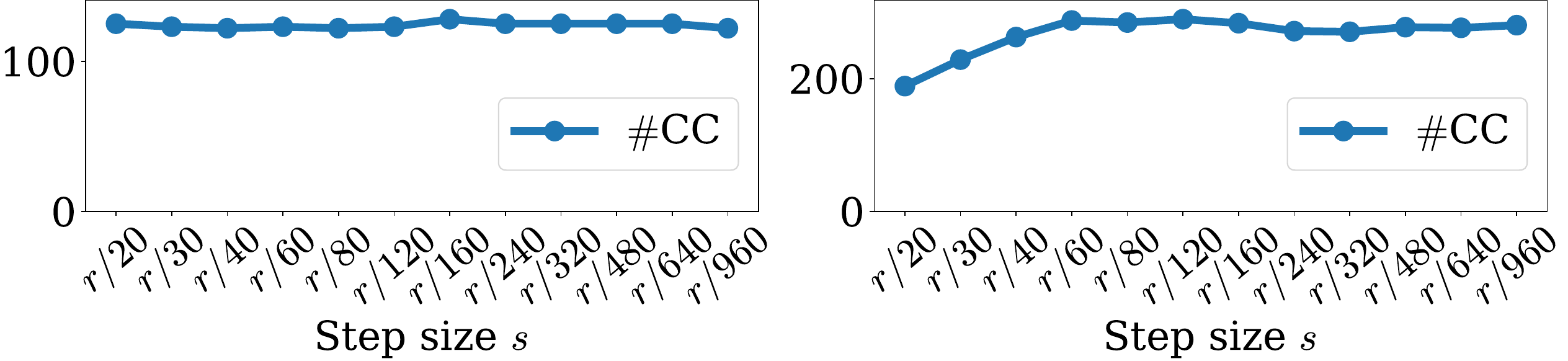}
\vspace{-6mm}
\caption{{\#CC} vs. step size $s$ on INR models: 2D Vortex Street (left) and 2D Heated Cylinder (right).}
\label{fig:inr_step_size_selection}
\vspace{-6mm}
\end{figure}

For the two INR models considered---Vortex Street INR and Heated Cylinder INR---we select the step size based on the stability of the number of connected components (\#CC), as illustrated in \cref{fig:inr_step_size_selection}. For both models, we select $s = r/80$.  

\subsection{Ablation Study on the Number of Seeds}
\begin{figure}[!ht]
\centering
\includegraphics[width=\linewidth]{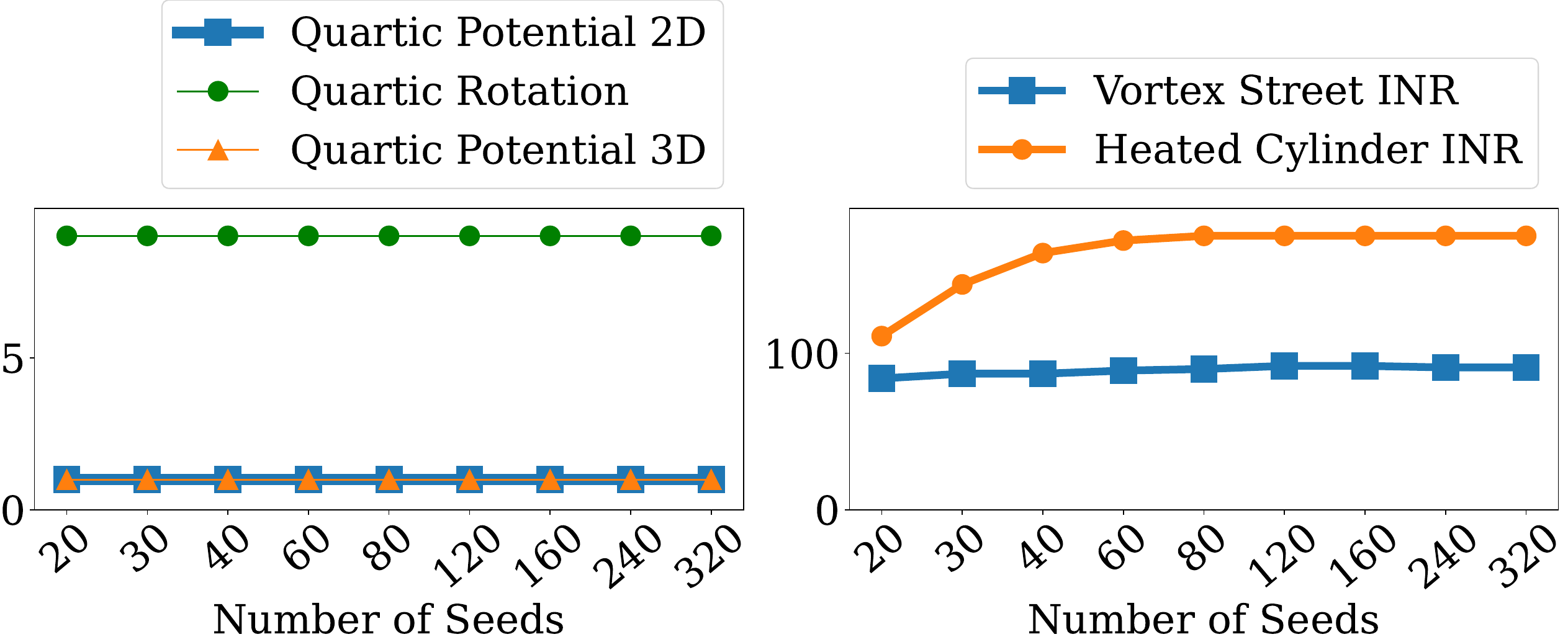}
\vspace{-6mm}
\caption{Number of initial points vs. the number of seeds along the shortest domain dimension when $t=0$ for analytic functions (left) and INR models (right).}
\label{fig:start_points_with_seeds}
\end{figure}

We report the number of initial points obtained from different numbers of seeds at $t=0$ in \cref{fig:start_points_with_seeds}. The analytic functions are robust to the choice of seed count. In our experiments, we sample 40 seeds along the shortest domain dimension and scale the sampling density proportionally along the remaining dimensions. For INR models, we instead use 80 samples once the results become stable.

\subsection{Ablation Study on $\epsilon$}
\begin{figure}[!ht]
\centering
\includegraphics[width=\linewidth]{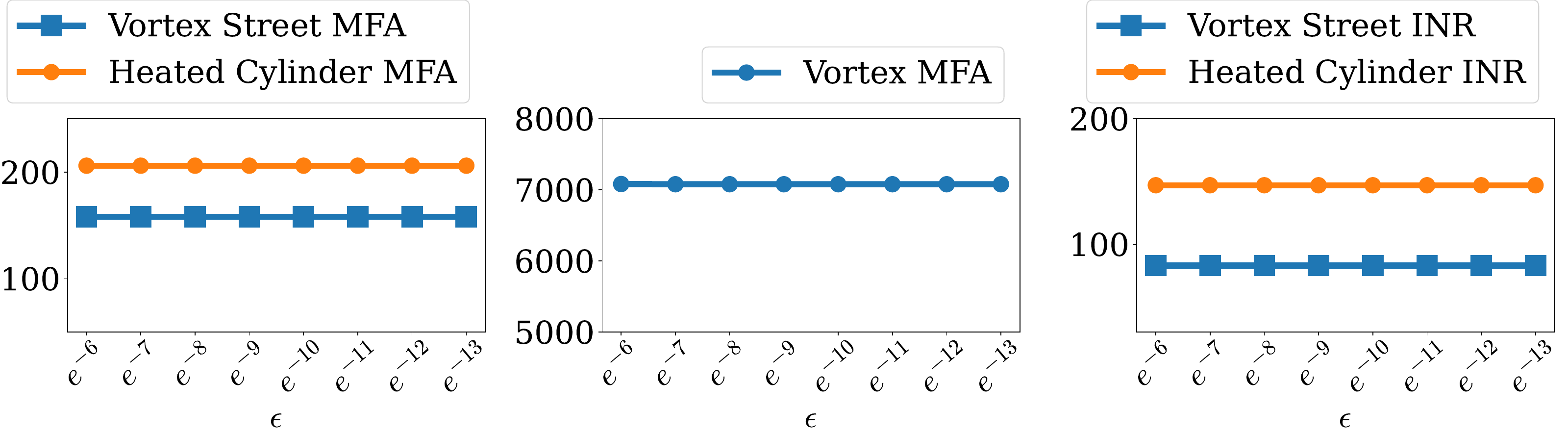}
\vspace{-6mm}
\caption{Number of initial points vs. $\epsilon$ when $t=0$ for different models.}
\label{fig:start_points_with_epsilon}
\end{figure}

We study the effect of $\epsilon$, which defines the threshold $||\nabla f_{\mathbf{x}}||<\epsilon$, on the number of initial points at $t=0$, as shown in \cref{fig:start_points_with_epsilon}. The results remain largely stable across a wide range of $\epsilon$, indicating low sensitivity to this parameter. Based on this observation, we choose $\epsilon=e^{-10}$ for all experiments.

\subsection{Ablation Study on $\epsilon_k$}
\begin{figure}[!ht]
\vspace{-2mm}
\centering
\includegraphics[width=\linewidth]{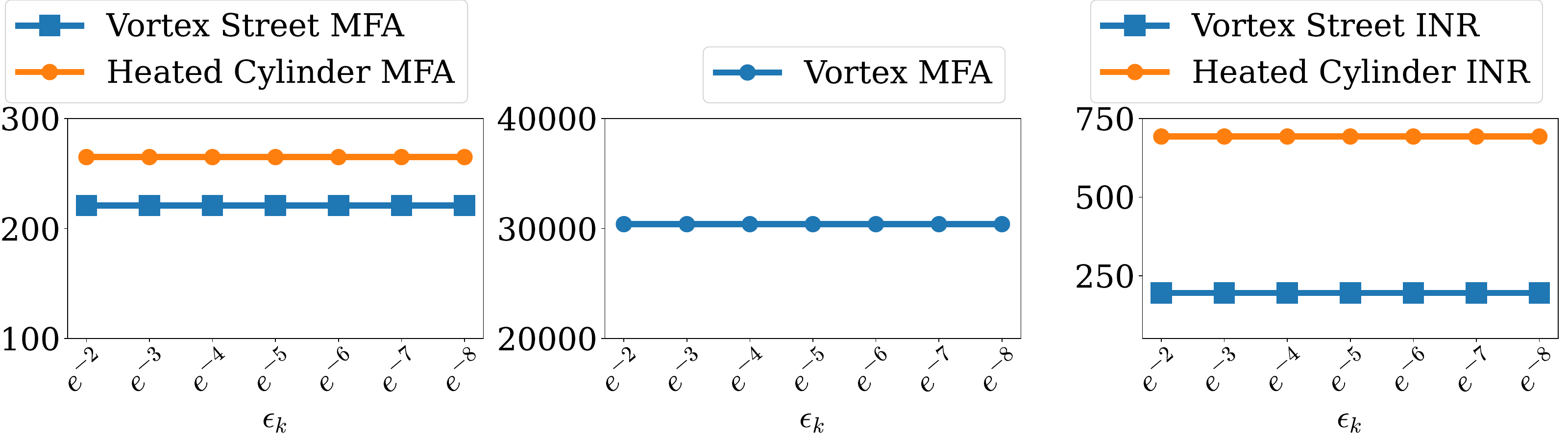}
\vspace{-6mm}
\caption{Number of degenerate points vs. $\epsilon_k$ for different models.}
\label{fig:degenerate_points_with_epsilon_k}
\end{figure}

We use $\epsilon_k$ as a threshold on the scale-normalized Hessian, i.e.,
$\frac{|\det(H_{\mathbf{x}}f)|}{\|H_{\mathbf{x}}f\|_F^{\,d}}<\epsilon_k$.
As shown in \cref{fig:degenerate_points_with_epsilon_k}, the number of degenerate points is largely insensitive to $\epsilon_k$. Based on this observation, we choose $\epsilon_k=e^{-6}$ for all experiments.

\subsection{Vortex Street MFA Model}
\begin{figure}[!ht]
\centering
\vspace{-1mm}
\includegraphics[width=\linewidth]{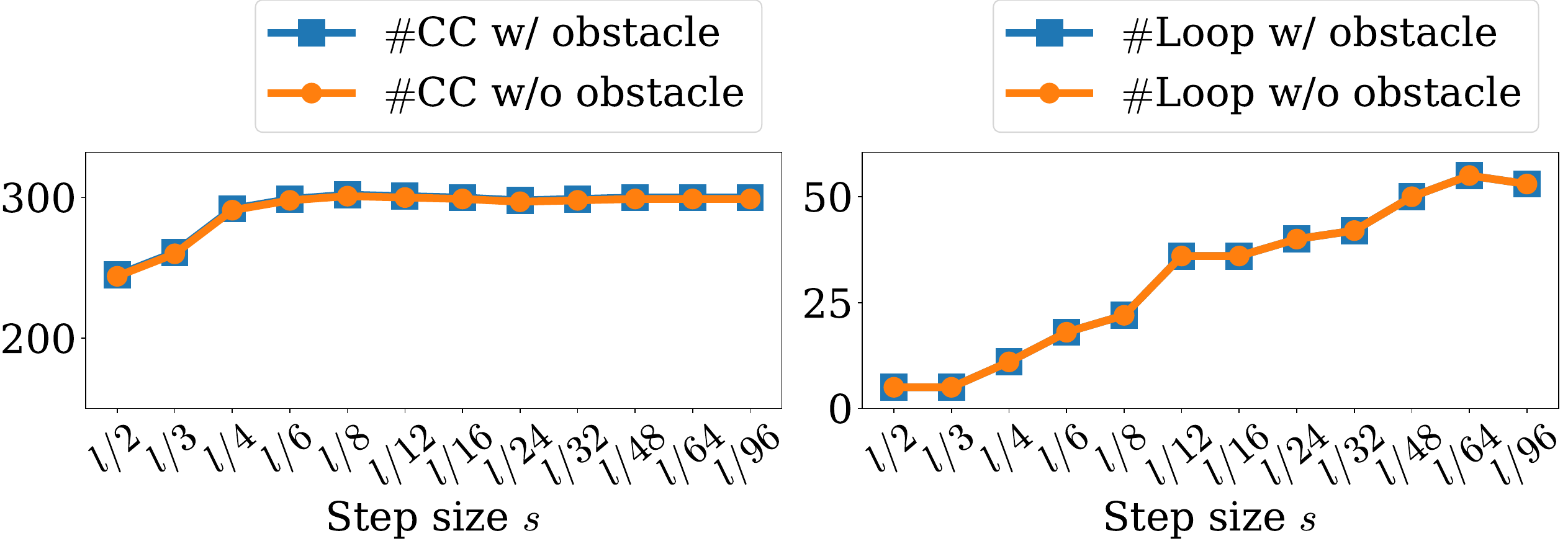}
\vspace{-6mm}
\caption{Vortex Street MFA model. Left: number of connected components vs. step size $s$. Right: number of loops vs. step size $s$.}
\label{fig:vortex_street_mfa_step_size_selection}
\vspace{-3mm}
\end{figure}

In \cref{fig:vortex_street_mfa_step_size_selection}, the number of connected components stabilizes as the step size decreases, indicating that the primary structure of the critical-point trajectories is recovered once the step size is sufficiently small. Because this metric captures large-scale structure and converges reliably, we use it to select the step size.

The number of loops exhibits different behavior. As the step size decreases, previously merged or marginal cycles may separate, leading to additional loops even after the number of connected components has largely stabilized.

\begin{figure}[!ht]
\centering
\vspace{-1mm}
\includegraphics[width=0.5\linewidth]{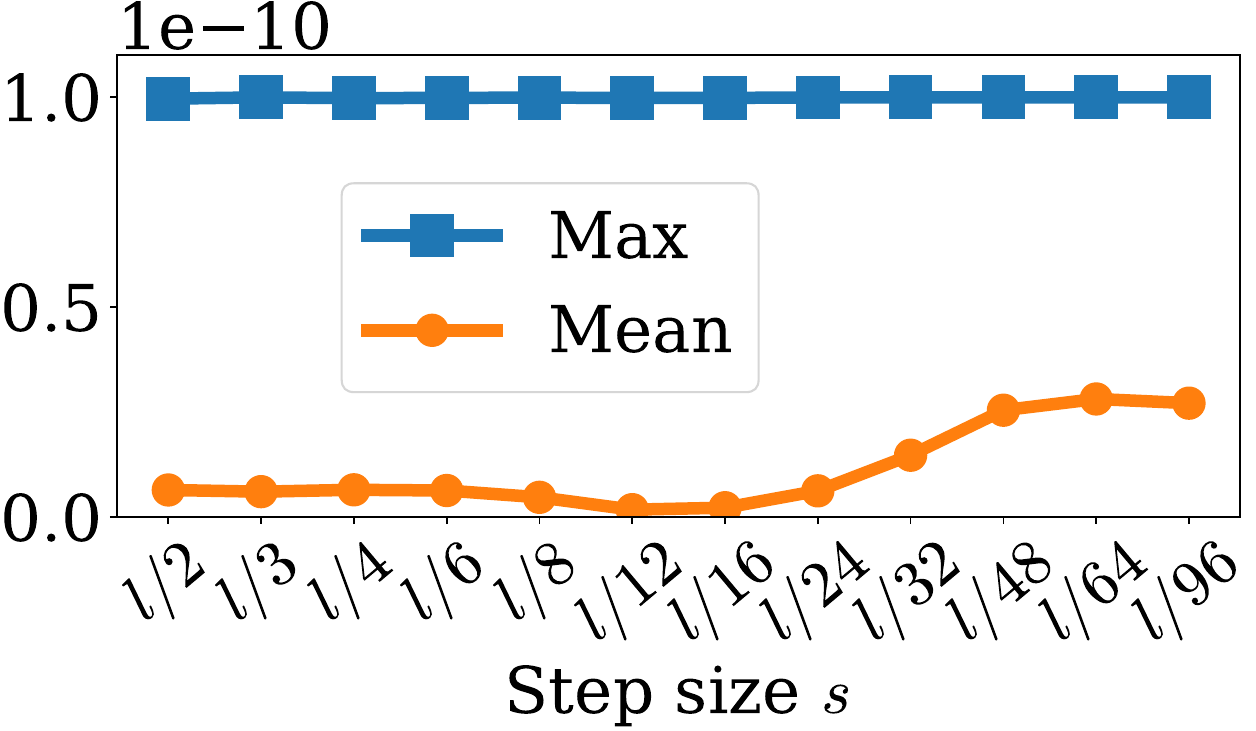}
\vspace{-2mm}
\caption{Vortex Street MFA model: spatial gradient norm vs. step size $s$.}
\label{fig:vortex_street_mfa_grad_norm}
\vspace{-3mm}
\end{figure}

\cref{fig:vortex_street_mfa_grad_norm} shows that the gradient norm remains below $e^{-10}$ for all step sizes, indicating stable implicit integration.

\subsection{Heated Cylinder MFA Model}
\begin{figure}[!ht]
\centering
\vspace{-1mm}
\includegraphics[width=\linewidth]{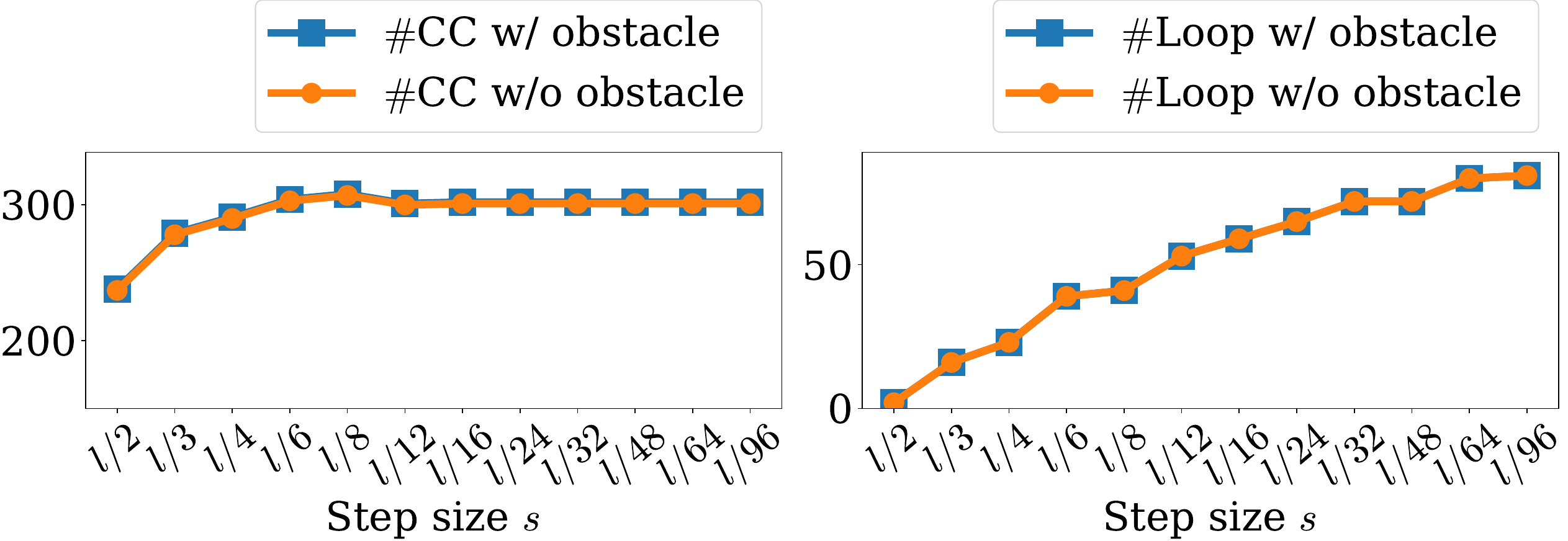}
\vspace{-6mm}
\caption{Heated Cylinder MFA model. Left: number of connected components vs. step size $s$. Right: number of loops vs. step size $s$.}
\label{fig:heated_cylinder_mfa_step_size_selection}
\vspace{-3mm}
\end{figure}

\begin{figure}[!ht]
\centering
\vspace{-1mm}
\includegraphics[width=0.5\linewidth]{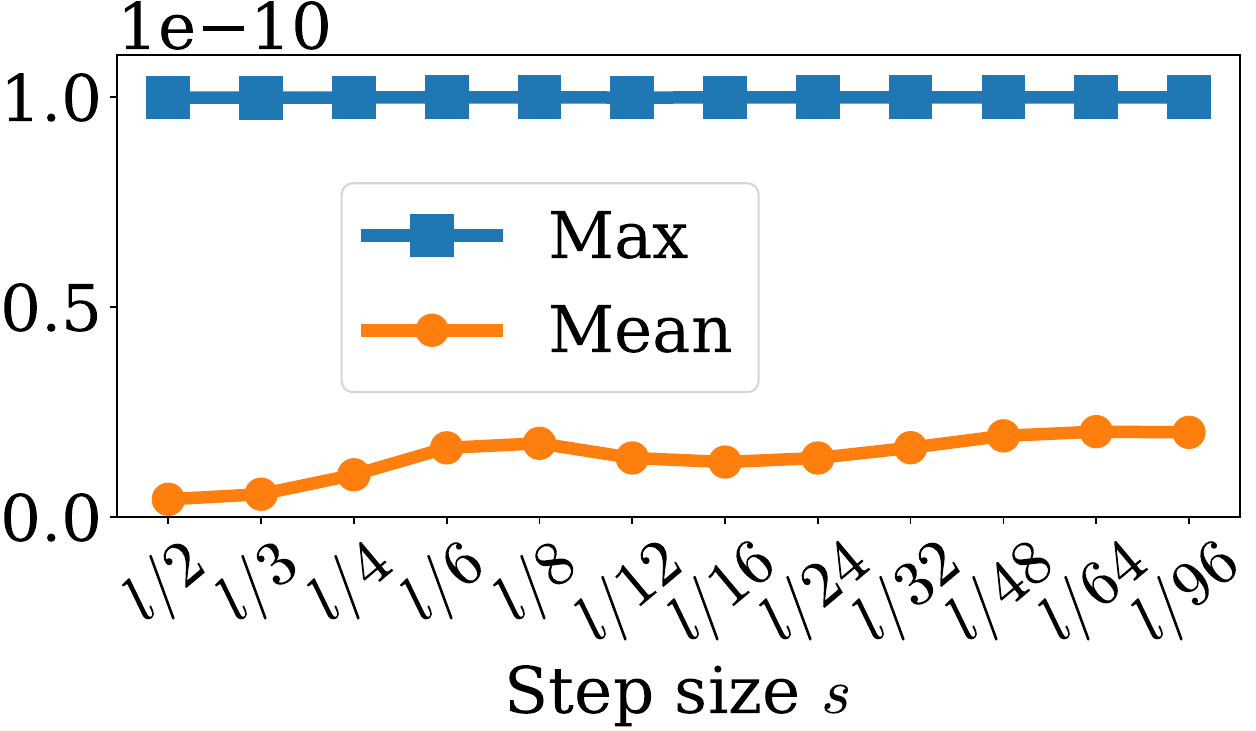}
\vspace{-2mm}
\caption{Heated Cylinder MFA model: spatial gradient norm vs. step size $s$.}
\label{fig:heated_cylinder_mfa_grad_norm}
\vspace{-4mm}
\end{figure}

A similar pattern is observed in \cref{fig:heated_cylinder_mfa_step_size_selection}. The number of connected components converges as the step size decreases, indicating that large-scale flow regions are reliably captured even at moderate step sizes. In contrast, the number of loops exhibits fluctuations, as smaller step sizes separate near-touching trajectories and reveal additional short-lived cycles.

As shown in \cref{fig:heated_cylinder_mfa_grad_norm}, the spatial gradient norm remains below $e^{-10}$ across all step sizes, confirming stable integration.

\subsection{Vortex MFA Model}
\begin{figure}[!ht]
\centering
\vspace{-1mm}
\includegraphics[width=\linewidth]{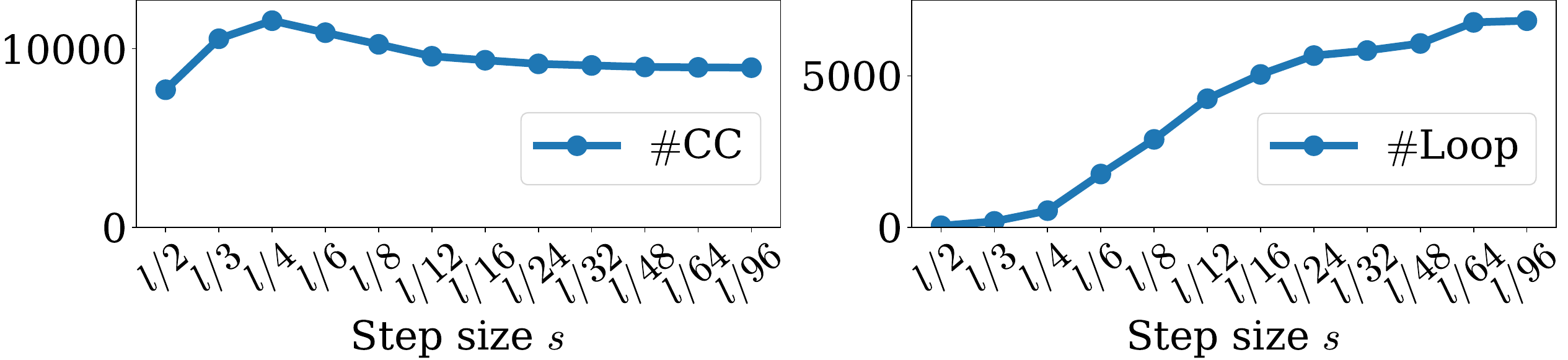}
\vspace{-6mm}
\caption{Vortex MFA model. Left: number of connected components vs. step size $s$. Right: number of loops vs. step size $s$.}
\label{fig:vortex_mfa_step_size_selection}
\vspace{-3mm}
\end{figure}

\begin{figure}[!ht]
\centering
\vspace{-1mm}
\includegraphics[width=0.5\linewidth]{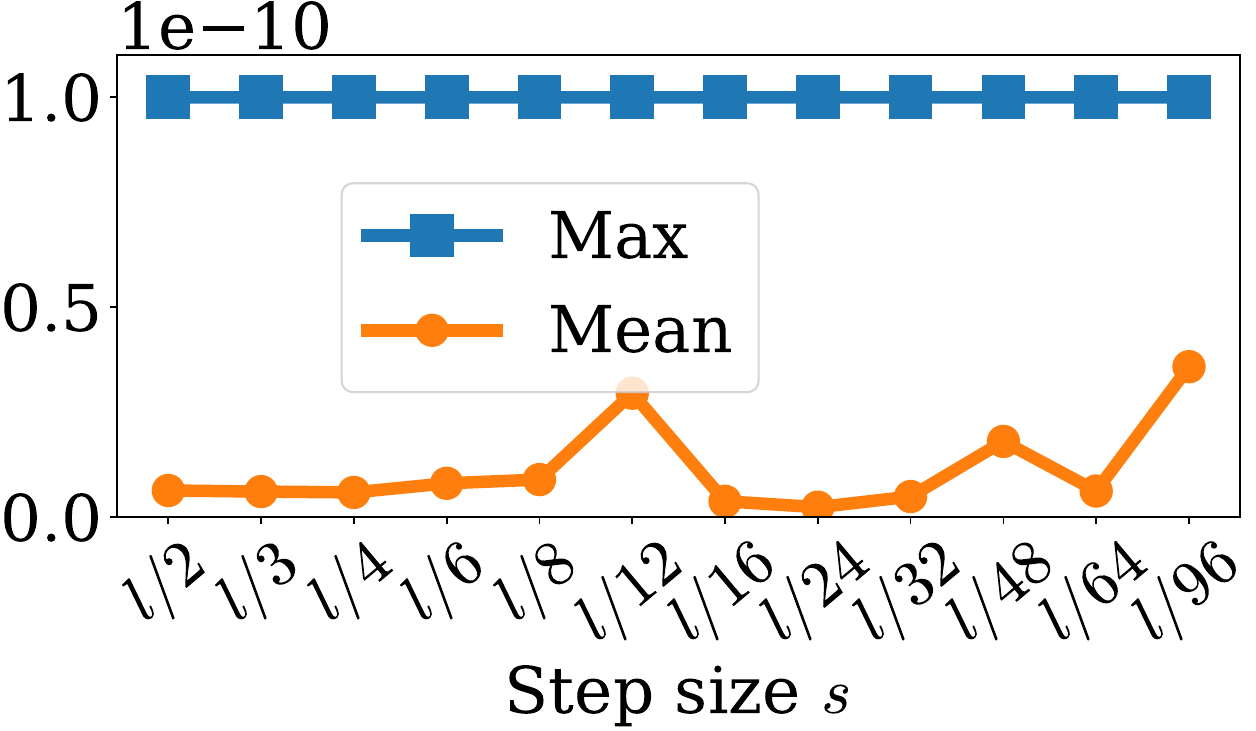}
\vspace{-2mm}
\caption{Vortex MFA model: spatial gradient norm vs. step size $s$.}
\label{fig:vortex_mfa_grad_norm}
\vspace{-3mm}
\end{figure}
In \cref{fig:vortex_mfa_step_size_selection}, the number of connected components converges as the step size decreases, indicating that large-scale flow regions are reliably captured even at moderate step sizes. As the step size decreases, previously merged or marginal cycles begin to separate, leading to the emergence of new loops even after the number of connected components has largely stabilized. As shown in \cref{fig:vortex_mfa_grad_norm}, the spatial gradient norm remains below $e^{-10}$ across all step sizes, confirming stable integration.

\subsection{Vortex Street INR Model}
\begin{figure}[!ht]
\centering
\vspace{-1mm}
\includegraphics[width=\linewidth]{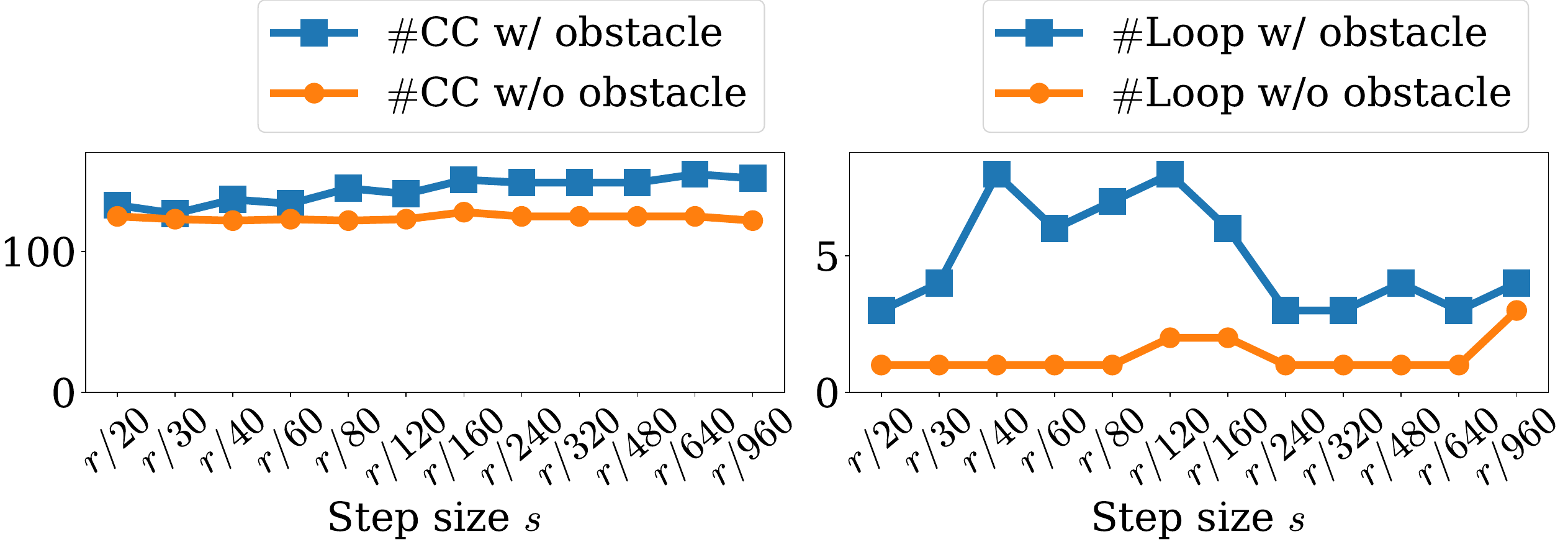}
\vspace{-6mm}
\caption{Vortex Street INR model. Left: number of connected components vs. step size $s$. Right: number of loops vs. step size $s$.}
\label{fig:vortex_street_inr_step_size_selection}
\vspace{-3mm}
\end{figure}

\begin{figure}[!ht]
\centering
\vspace{-1mm}
\includegraphics[width=0.5\linewidth]{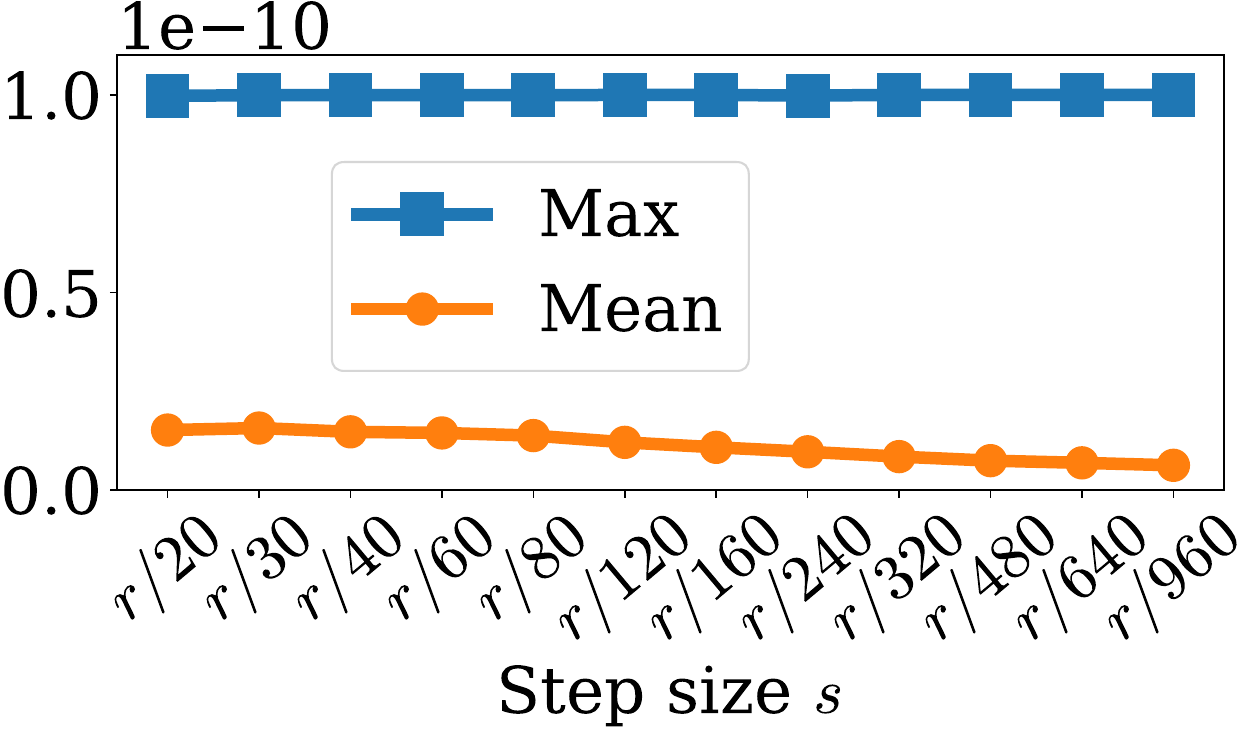}
\vspace{-2mm}
\caption{Vortex Street INR model: spatial gradient norm vs. step size $s$.}
\label{fig:vortex_street_inr_grad_norm}
\vspace{-3mm}
\end{figure}

In \cref{fig:vortex_street_inr_step_size_selection}, the number of connected components remains relatively stable across step sizes, suggesting that the dominant flow regions are consistently captured. The loop count also remains small, ranging only between 1 and 3 in the obstacle-free setting.

In \cref{fig:vortex_street_inr_grad_norm}, the spatial gradient norm remains below $e^{-10}$ across all step sizes, confirming stable integration.

\subsection{Heated Cylinder INR Model}
\begin{figure}[!ht]
\centering
\vspace{-2mm}
\includegraphics[width=\linewidth]{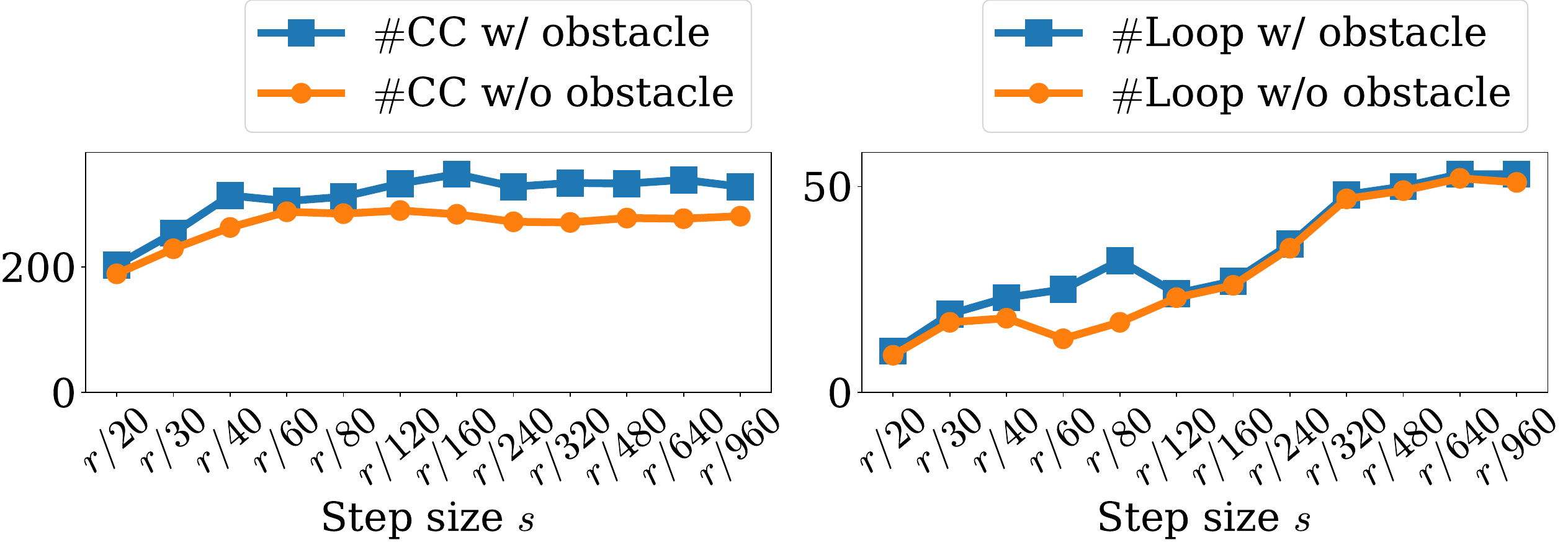}
\vspace{-6mm}
\caption{Heated Cylinder INR model. Left: number of connected components vs. step size $s$. Right: number of loops vs. step size $s$.}
\label{fig:heated_cylinder_inr_step_size_selection}
\vspace{-3mm}
\end{figure}

\begin{figure}[!ht]
\centering
\includegraphics[width=0.5\linewidth]{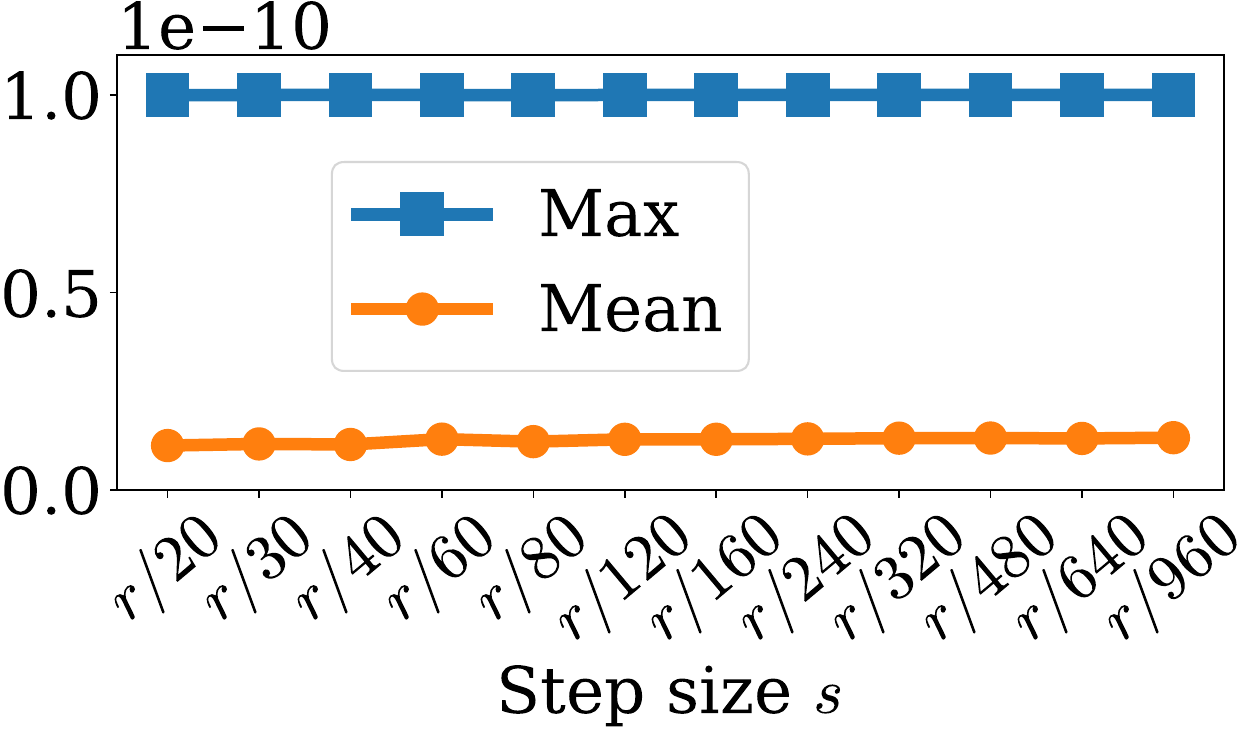}
\vspace{-2mm}
\caption{Heated Cylinder INR model: spatial gradient norm vs. step size $s$.}
\label{fig:heated_cylinder_inr_grad_norm}
\vspace{-3mm}
\end{figure}

For the Heated Cylinder INR model, the number of loops does not stabilize until very small step sizes (see \cref{fig:heated_cylinder_inr_step_size_selection}, right). As the step size decreases, previously merged or marginal cycles become distinguishable, leading to the emergence of additional loops even after the number of connected components has largely stabilized. In contrast, the number of connected components gradually stabilizes, indicating that the primary structure is reliably captured.

As shown in \cref{fig:heated_cylinder_inr_grad_norm}, the spatial gradient norm remains below $e^{-10}$ across all step sizes, demonstrating stable integration.

%% file: sec-app-necessity-of-newton-correction.tex
\section{Necessity of Newton Correction}
\label{sec:necessity-of-newton-correction}

We quantify the effect of Newton correction by comparing implicit extraction with and without correction on the Heated Cylinder MFA model. As reported in \cref{tab:gradient-magnitude-newton-correction}, Newton correction reduces the mean spatial gradient norm from $6.43 \times 10^{-2}$ to $1.30 \times 10^{-11}$ and the maximum norm from $2.58$ to $1.00 \times 10^{-10}$. These reductions correspond to factors of approximately $4.95 \times 10^{9}$ and $2.58 \times 10^{10}$, respectively, or roughly ten orders of magnitude.

\begin{table}[!ht]
\centering
\small
\vspace{-2mm}
\begin{tabular}{lcc}
\hline
Statistic & With correction & Without correction \\
\hline
Mean    & $1.30 \times 10^{-11}$ & $6.43 \times 10^{-2}$ \\
Maximum & $1.00 \times 10^{-10}$ & $2.58$ \\
\hline
\end{tabular}
\vspace{-2mm}
\caption{Mean and maximum spatial gradient norms, $\|\nabla_{\mathbf{x}}f\|$, over all extracted trajectory points of the Heated Cylinder MFA model, with and without Newton correction.}
\label{tab:gradient-magnitude-newton-correction}
\vspace{-3mm}
\end{table}

The maximum residual after correction matches the prescribed tolerance, $\epsilon=10^{-10}$, whereas the uncorrected residuals exceed this tolerance by many orders of magnitude. This difference highlights the role of Newton correction in enforcing the critical-point condition, $\nabla_{\mathbf{x}}f(\mathbf{x},t)=\mathbf{0}$. RK4 updates can accumulate numerical drift away from this condition, while Newton correction refines the predicted positions to satisfy it within the prescribed tolerance. Thus, RK4 tracing alone does not provide sufficient accuracy in this experiment.

\begin{figure*}
\centering
\vspace{-3mm}
\includegraphics[width=\linewidth]{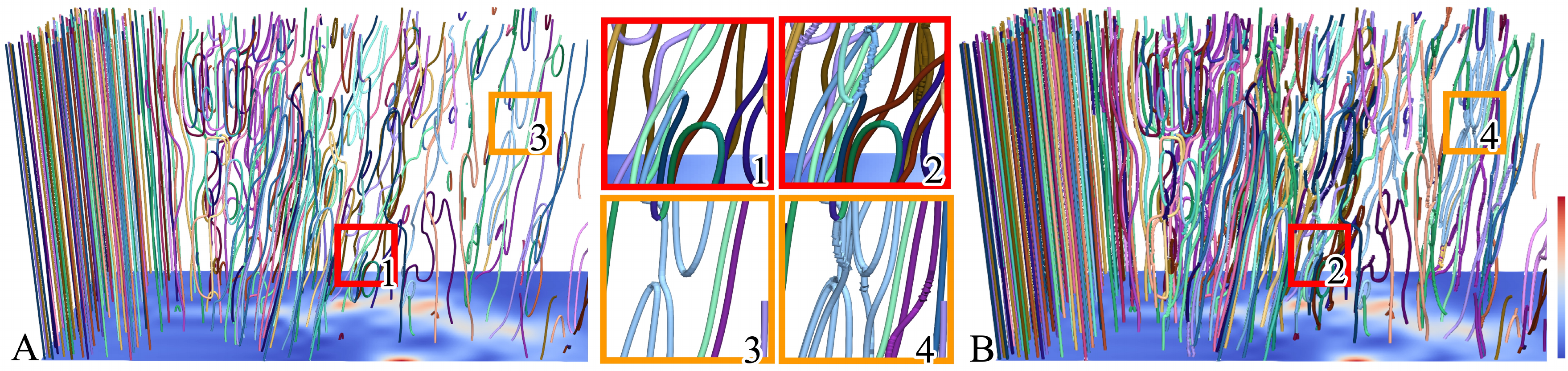}
\vspace{-6mm}
\caption{Critical-point trajectories of the Heated Cylinder MFA model extracted by implicit extraction with Newton correction (A) and without Newton correction (B). The central panels enlarge the regions highlighted by the colored boxes: (1) and (3) correspond to (A), while (2) and (4) correspond to (B).}
\label{fig:tracking-without-newton}
\vspace{-4mm}
\end{figure*}

The visual comparison in \cref{fig:tracking-without-newton} supports these numerical results. Without correction, panels (2) and (4) show additional trajectories compared with the corrected results in panels (1) and (3). When the same underlying trajectory is traced from different seeds, RK4 integration errors accumulate differently along each trace. Without Newton correction, these traces can drift far enough apart to no longer satisfy the spatial--temporal proximity criteria used for duplicate removal (\cref{sec:particle-tracing}). Consequently, deduplication retains them as distinct trajectories even though they represent the same underlying trajectory. Newton correction suppresses this drift, helping to maintain the critical-point condition and prevent redundant traces from surviving deduplication.

%% file: sec-app-necessity-of-degenerate-cases.tex
\section{Necessity of Explicit Degenerate-Case Handling}
\label{sec:necessity-of-degenerate-cases}

Explicit handling of degenerate cases is necessary to recover branching events in our continuation framework. The velocity in \cref{eq:velocity} is defined only when the spatial Hessian is nonsingular. At a degenerate event, following a single trajectory neither identifies all incident branches nor establishes their connectivity. For example, in the 2D Quartic Potential model, the central trajectory $(0,0,t)$ remains a valid critical-point trajectory across $t=1$, while two additional branches, $(\pm\sqrt{t-1},0,t)$, emerge for $1<t<\pi/2$. Thus, continuation along a single trajectory can satisfy the zero-gradient condition while missing the emerging branches.

\begin{figure*}[!ht]
\centering
\includegraphics[width=\linewidth]{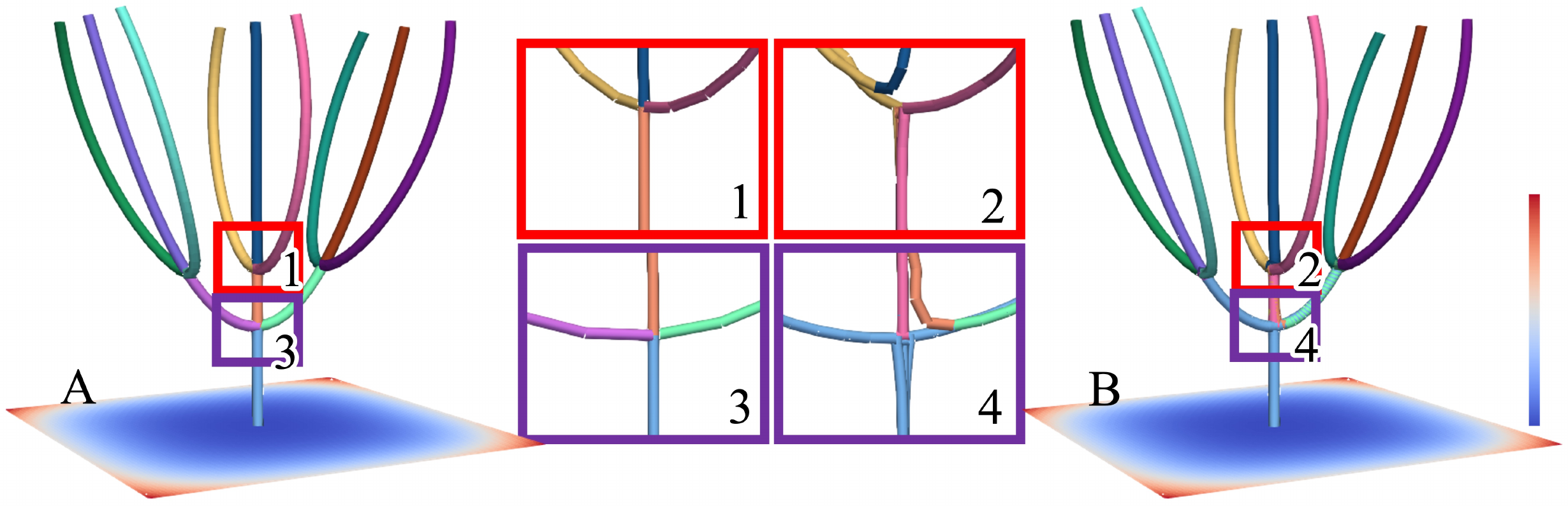}
\vspace{-6mm}
\caption{Critical-point trajectories of the 2D Quartic Potential model extracted by implicit extraction (A) and SFFF (B). The central panels enlarge the regions highlighted by the colored boxes: (1) and (3) correspond to (A), while (2) and (4) correspond to (B).}
\label{fig:quartic-potential-2d-fff}
\vspace{-4mm}
\end{figure*}

\cref{fig:quartic-potential-2d-fff} compares our method with SFFF on this model, which contains four known degenerate points. The enlarged views in panels (1) and (3) show that our method recovers the splitting events and connects the incident branches at their shared degenerate points. In contrast, the corresponding SFFF results in panels (2) and (4) do not reproduce the correct branch junctions. This comparison illustrates that tracking regular trajectory segments alone is insufficient to recover the branching structure.

%% file: sec-app-analytic-validation.tex
\section{Analytic Reference Validation}
\label{sec:analytic-reference-validation}

We evaluate the three analytic examples using the selected step size $s=r/40=0.1$. All comparisons use the original, unnormalized coordinates over the full time interval $[0,4]$. We evaluate the recovered graphs using their stored vertices and straight edges, and construct the references from the analytic critical-point expressions in \cref{sec:configuration}.

\para{Hausdorff distance and component correspondence.}
For compact sets $A,B\subset\mathbb{R}^{d+1}$, the symmetric Hausdorff distance is
\begin{equation}
\begin{split}
 H(A,B)&=\max\{h(A,B),h(B,A)\},\\
 h(A,B)&=\sup_{a\in A} \inf_{b\in B}\|a-b\|_2.
\end{split}
\label{eq:analytic-hausdorff}
\end{equation}
The coordinates are $(x,y,t)$ for the 2D examples and $(x,y,z,t)$ for the 3D example. The nearest-point search is not restricted to points at the same time. Each quartic-potential example has a single connected reference graph, so we compare the complete graphs. For Quartic Rotation, we match the nine recovered connected components to the nine analytic trajectories, including the stationary central trajectory. We verify that every vertex is consistent with its assigned component and report $\max_j H(G_j,C_j)$, where $G_j$ and $C_j$ are corresponding recovered and analytic components. We do not subdivide components into smaller branches for this calculation.

\para{Reference sampling.}
We densely sample the analytic trajectories and connect consecutive samples with straight segments. We select the sampling density so that the maximum interpolation error between each analytic curve and its reference polyline is at most $\varepsilon_{\rm ref}=2.5 \times 10^{-9}$.

We choose this reference tolerance to ensure that the final bounds on the Hausdorff distance form an interval of width at most $10^{-8}$. The adaptive polyline comparison described below produces an interval of width at most $5 \times 10^{-9}$. Expanding the interval by $\varepsilon_{\rm ref}$ at each end to account for reference interpolation yields a total width of at most $5 \times 10^{-9}+2\varepsilon_{\rm ref}=10^{-8}$.

We sample the rotating trajectories uniformly in time. For quartic-potential pieces that emerge at a degenerate event, we sample uniformly in the square root of the elapsed time since the event, keeping the derivatives with respect to the sampling parameter bounded. Sampling covers the full interval of each piece, including its event endpoints.

In our experiments, we use $301{,}734$ reference samples for 2D Quartic Potential, $247{,}694$ for 2D Quartic Rotation, and $1{,}202{,}522$ for 3D Quartic Potential. These totals sum the samples across all analytic curve pieces, counting shared event endpoints separately for each piece.

\para{Distances between polylines.}
We compute the distance from a point to the other polyline as the minimum Euclidean distance to its line segments, considering both their endpoints and interiors. We estimate the maximum of this distance along the source polyline by adaptively subdividing its segments. We then repeat the calculation in the reverse direction and take the larger of the two directional maxima to estimate the symmetric Hausdorff distance in \cref{eq:analytic-hausdorff}.

\para{Spatial vertex error at identical times.}
For a recovered vertex $p_i=(\mathbf{x}_i,t_i)$, let $C_i(t_i)$ denote the set of analytic spatial roots at time $t_i$, restricted to the corresponding component for Quartic Rotation. We report
\begin{equation}
 E_V=\max_i e_i,\qquad
 e_i=\min_{\mathbf{y}\in C_i(t_i)}\|\mathbf{x}_i-\mathbf{y}\|_2.
\label{eq:analytic-vertex-error}
\end{equation}
This measure assesses vertex localization at identical times, whereas the Hausdorff distance also captures deviations of straight edges from the curved analytic trajectories.

%% file: sec-app-real-data-reference.tex
\section{Local Reference Validation for Scientific Models}
\label{sec:INR-MFA-validation}

We provide local reference comparisons for the scientific models in \cref{sec:scientific-results}. Since ground-truth trajectories are unavailable for these models, we compare implicit extraction with LWM results obtained from denser samples of the same continuous models.

These densely sampled LWM results remain discrete approximations. Comparisons at two sampling resolutions therefore do not establish convergence or correctness of the complete trajectory graph.

\para{Reference construction.}
For each of the Vortex Street and Heated Cylinder MFA and INR models, we refine the three space--time regions highlighted in the corresponding figure (\cref{fig:vortex-street-mfa,fig:heated-cylinder-mfa,fig:vortex-street-inr,fig:heated-cylinder-inr}). We evaluate each continuous model directly on a grid with tenfold resolution in $x$, $y$, and $t$, increasing the sample count by approximately $10^3$. We then apply LWM to the resulting scalar fields and compare the extracted trajectories with those from implicit extraction and LWM at the original sampling resolution. Refining the entire Vortex Street MFA domain by the same factor would increase the sampled data size from approximately 430~MB to 430~GB, before accounting for tracking overhead. Local refinement makes these reference comparisons feasible.

\begin{figure*}[p]
\centering
\includegraphics[width=\linewidth]{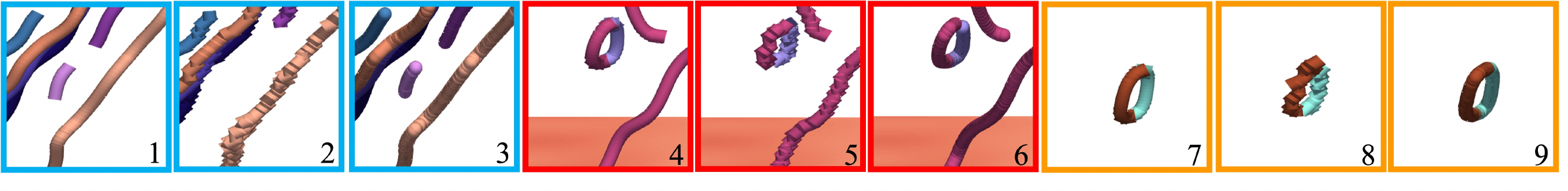}
\vspace{-6mm}
\caption{Local reference comparison for the Vortex Street MFA model in the three regions highlighted in \cref{fig:vortex-street-mfa}. Panels (1,4,7): implicit extraction; (2,5,8): LWM at the original sampling resolution; (3,6,9): LWM after tenfold refinement in each of $x$, $y$, and $t$.}
\label{fig:vortex-street-mfa-upsample-ttk}
\end{figure*}

\begin{figure*}[p]
\centering
\includegraphics[width=\linewidth]{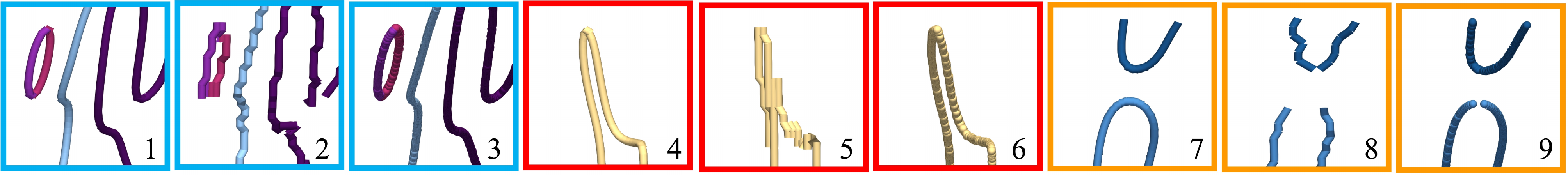}
\vspace{-6mm}
\caption{Local reference comparison for the Heated Cylinder MFA model in the three regions highlighted in \cref{fig:heated-cylinder-mfa}. Panels (1,4,7): implicit extraction; (2,5,8): original-resolution LWM; (3,6,9): LWM after tenfold refinement in $x$, $y$, and $t$.}
\label{fig:heated-cylinder-mfa-upsample-ttk}
\end{figure*}

\para{MFA models.}
For Vortex Street in \cref{fig:vortex-street-mfa-upsample-ttk}, the refined LWM trajectories follow the implicit trajectories more closely and exhibit less pronounced grid-aligned zigzags. In particular, panel (9) recovers the closed loop shown in panel (7), whereas panel (6) misses the loop shown in panel (4). For Heated Cylinder in \cref{fig:heated-cylinder-mfa-upsample-ttk}, refined LWM recovers the small closed loop in panel (3), consistent with implicit extraction in panel (1), whereas the original-resolution LWM result in panel (2) leaves this loop open. The curved trajectories in panels (6) and (9) also agree more closely with their implicit counterparts in panels (4) and (7).

\begin{figure*}[p]
\centering
\includegraphics[width=\linewidth]{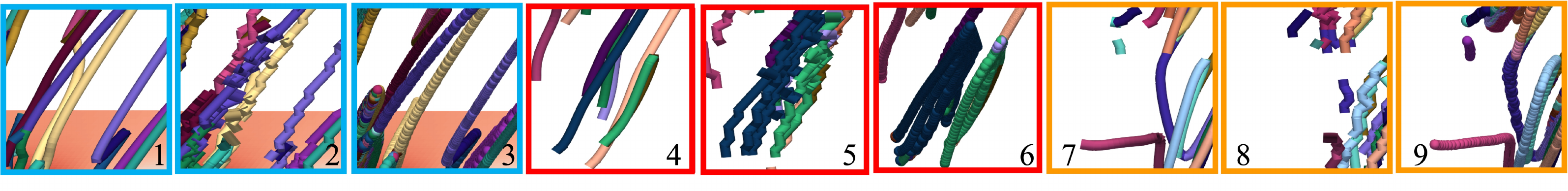}
\vspace{-6mm}
\caption{Local reference comparison for the Vortex Street INR model in the three regions highlighted in \cref{fig:vortex-street-inr}. Panels (1,4,7): implicit extraction; (2,5,8): original-resolution LWM; (3,6,9): LWM after tenfold refinement in $x$, $y$, and $t$.}
\label{fig:vortex-street-inr-upsample-ttk}
\end{figure*}

\begin{figure*}[p]
\centering
\includegraphics[width=\linewidth]{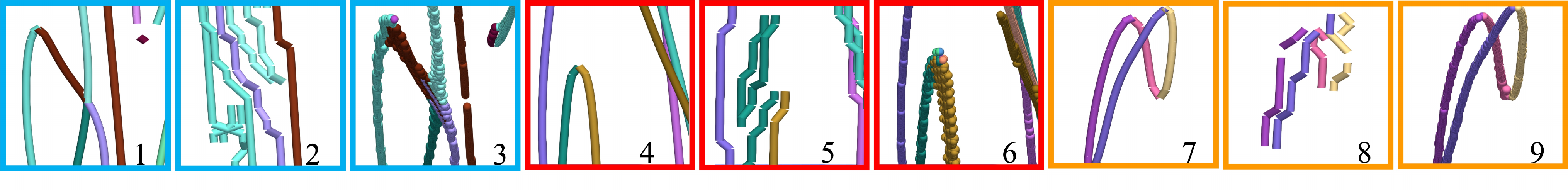}
\vspace{-6mm}
\caption{Local reference comparison for the Heated Cylinder INR model in the three regions highlighted in \cref{fig:heated-cylinder-inr}. Panels (1,4,7): implicit extraction; (2,5,8): original-resolution LWM; (3,6,9): LWM after tenfold refinement in $x$, $y$, and $t$.}
\label{fig:heated-cylinder-inr-upsample-ttk}
\end{figure*}

\para{INR models.}
For Vortex Street in \cref{fig:vortex-street-inr-upsample-ttk}, denser sampling reduces the visible zigzags in the LWM trajectories and improves their geometric agreement with implicit extraction. However, closely spaced traces persist in panel (6), so this comparison does not establish one-to-one branch correspondence in that region. For Heated Cylinder in \cref{fig:heated-cylinder-inr-upsample-ttk}, the densely sampled LWM reference trajectories in panels (3), (6), and (9) follow the corresponding implicit trajectories in panels (1), (4), and (7) more closely. Together, these comparisons provide visual support for the local trajectories recovered by our method.

%% file: sec-app-additional-results.tex
\section{Additional Experimental Results}
\label{sec:additional-results}

\begin{figure*}[!ht]
\centering
\includegraphics[width=\linewidth]{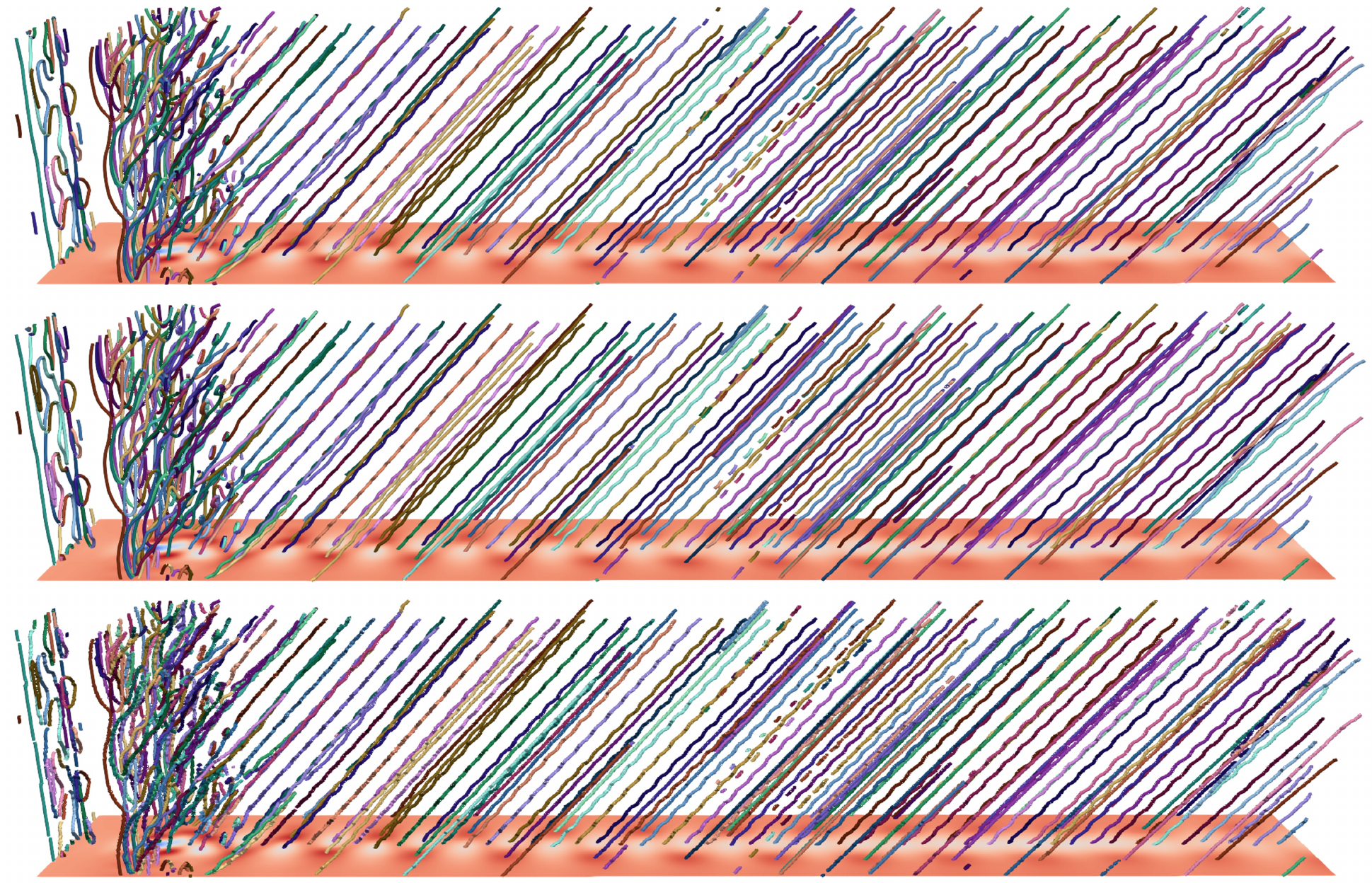}
\vspace{-5mm}
\caption{
Critical-point trajectories of the Vortex Street MFA model with obstacle, obtained via implicit extraction (top), SFFF (middle), and LWM (bottom).}
\label{fig:vortex-street-mfa-obstacle}
\end{figure*}

\begin{figure*}[!ht]
\centering
\includegraphics[width=\linewidth]{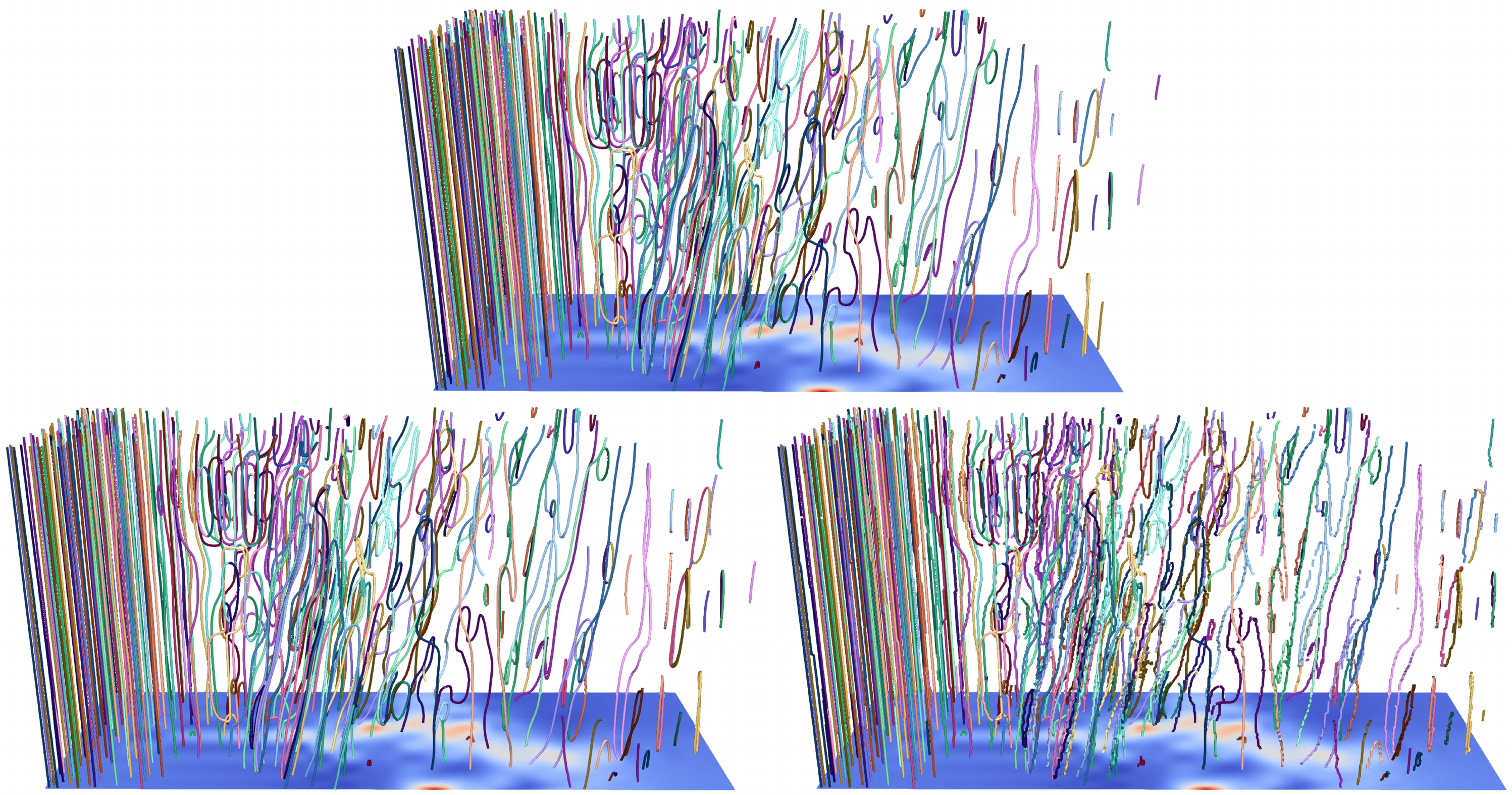}
\vspace{-5mm}
\caption{
Critical-point trajectories of the Heated Cylinder MFA model with obstacle, obtained via implicit extraction (top), SFFF (bottom left), and LWM (bottom right).}
\label{fig:heated-cylinder-mfa-obstacle}
\end{figure*}

\begin{figure*}[!ht]
\centering
\includegraphics[width=\linewidth]{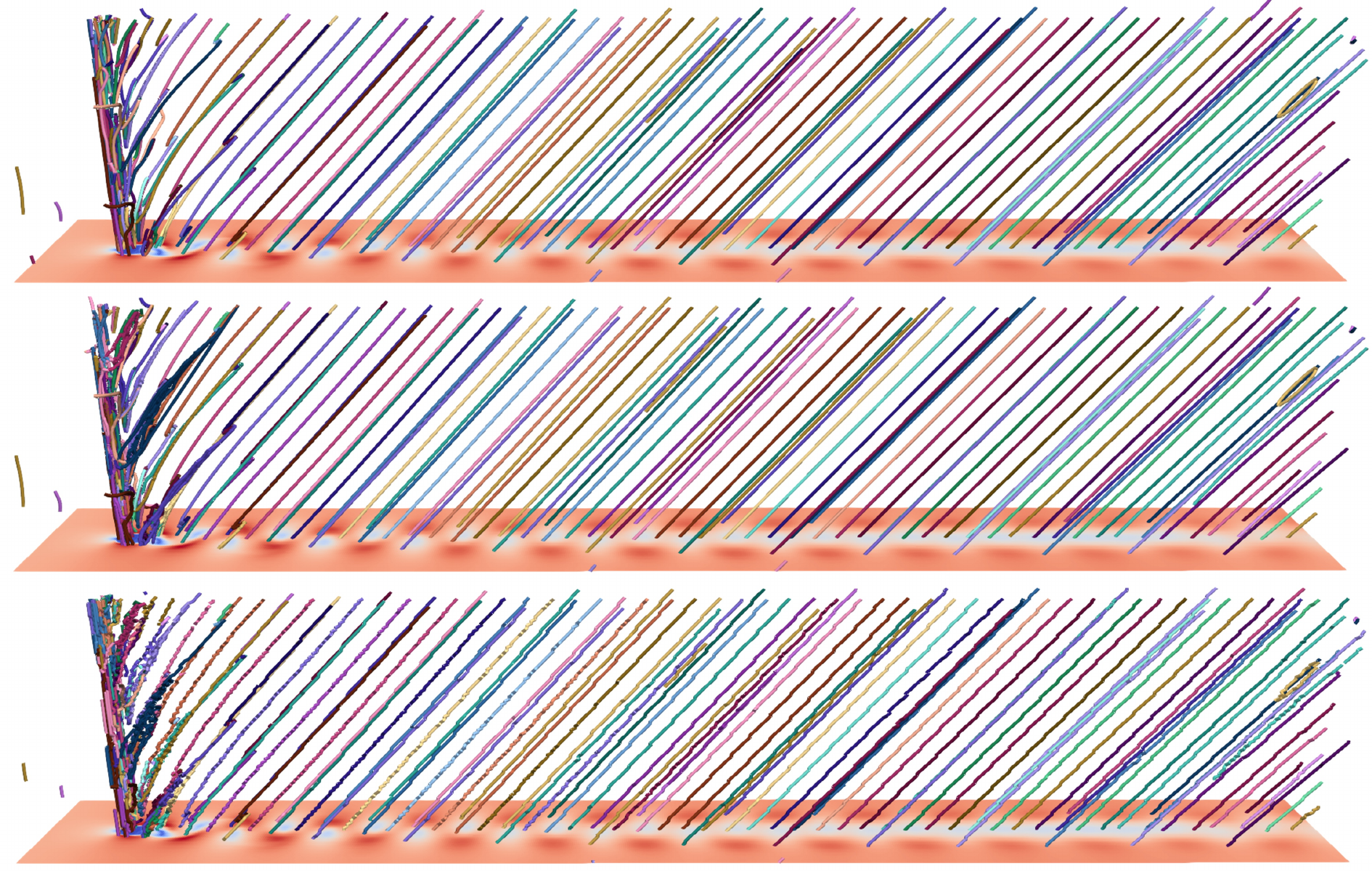}
\vspace{-5mm}
\caption{
Critical-point trajectories of the Vortex Street INR model with obstacle, obtained via implicit extraction (top), SFFF (middle), and LWM (bottom).}
\label{fig:vortex-street-inr-obstacle}
\end{figure*}

\begin{figure*}[!ht]
\centering
\includegraphics[width=\linewidth]{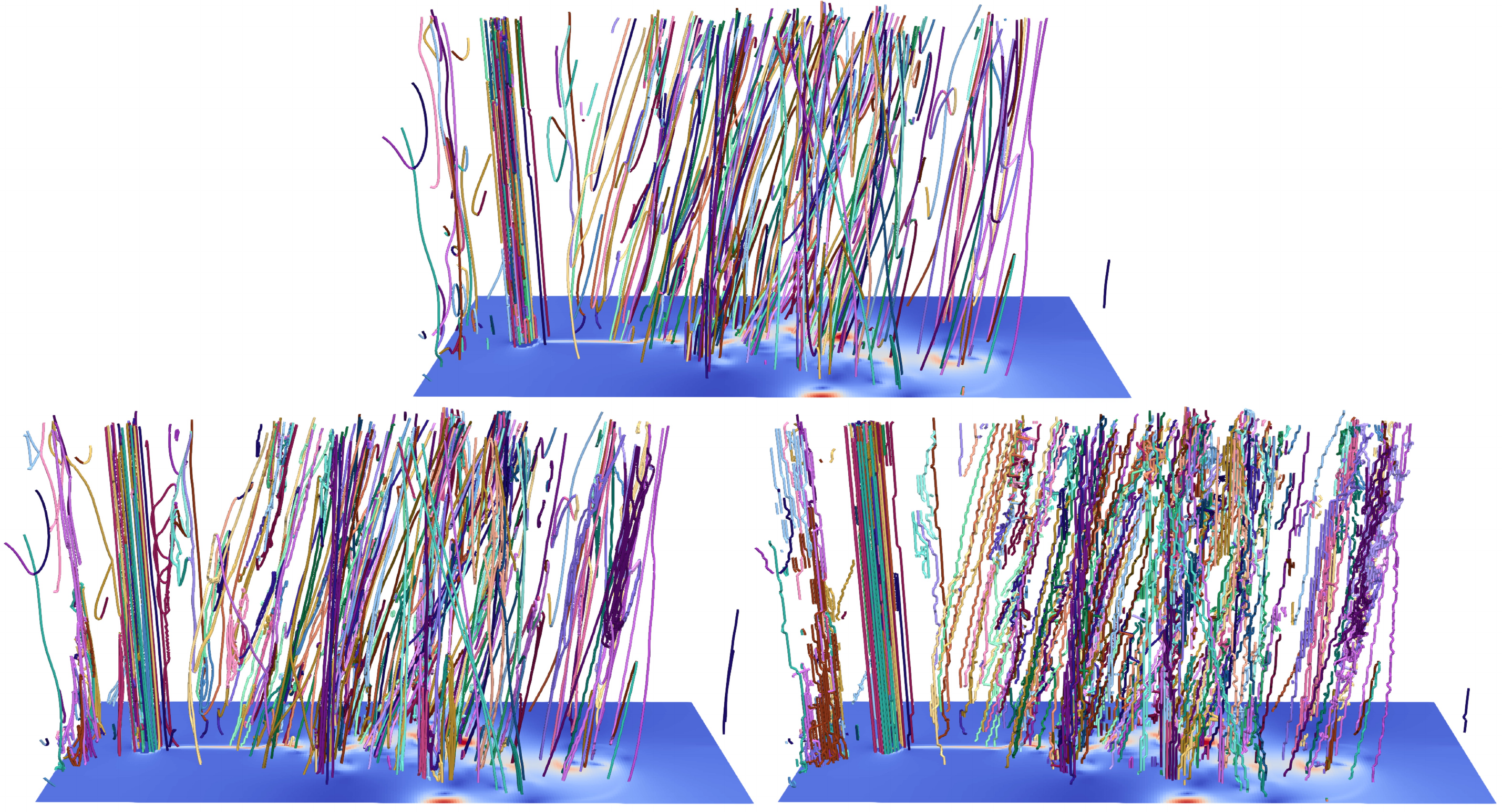}
\vspace{-5mm}
\caption{
Critical-point trajectories of the Heated Cylinder INR model with obstacle, obtained via implicit extraction (left), SFFF (bottom left), and LWM (bottom right).}
\label{fig:heated-cylinder-inr-obstacle}
\vspace{-3mm}
\end{figure*}

We additionally report results that retain the obstacle region, using the same model-specific step sizes. As shown in \cref{fig:vortex-street-mfa-obstacle,fig:heated-cylinder-mfa-obstacle,fig:vortex-street-inr-obstacle,fig:heated-cylinder-inr-obstacle}, implicit extraction recovers smooth trajectory structures that agree closely with those obtained by SFFF. In contrast, LWM produces visibly more zigzag trajectories. The dense vertical bands coincide with the cylindrical obstacle, where the velocity magnitude is identically zero. Consequently, the obstacle contains a non-isolated set of critical points, for which critical-point tracking is not meaningful. We therefore exclude the obstacle region and focus the analysis on the physically meaningful flow domain.

